\documentclass[letterpaper,twocolumn,showpacs,superscriptaddress]{quantumarticle}
\usepackage[caption=false]{subfig}
\usepackage{graphicx} 
\usepackage{epstopdf}
\usepackage{array}
\usepackage{verbatim}
\usepackage{amsmath,amsfonts,amssymb,amscd}
\usepackage{amsthm}
\usepackage{tabularx}
\usepackage{multirow}
\usepackage{stmaryrd}
\usepackage{enumerate}
\usepackage[ruled]{algorithm2e}
\usepackage{enumitem}
\usepackage{wasysym}
\usepackage{geometry}
\usepackage[numbers,sort&compress]{natbib}
\usepackage[colorlinks = true]{hyperref}

\theoremstyle{plain}

\theoremstyle{definition}

\newcommand{\eq}[1]{(\hyperref[eq:#1]{\ref*{eq:#1}})}

\renewcommand{\sec}[1]{\hyperref[sec:#1]{Section~\ref*{sec:#1}}}
\newcommand{\thrm}[1]{\hyperref[thm:#1]{Theorem~\ref*{thm:#1}}}
\newcommand{\lemm}[1]{\hyperref[lemm:#1]{Lemma~\ref*{lemm:#1}}}
\newcommand{\prop}[1]{\hyperref[prop:#1]{Proposition~\ref*{prop:#1}}}
\newcommand{\corr}[1]{\hyperref[corr:#1]{Corollary~\ref*{corr:#1}}}
\newcommand{\fig}[1]{\hyperref[fig:#1]{Figure~\ref*{fig:#1}}}

\newcommand{\ket}[1]{|#1\rangle}

\usepackage[usenames,dvipsnames]{color}
\usepackage{xcolor}

\DeclareMathAlphabet{\matheu}{U}{eus}{m}{n}

\newcolumntype{L}[1]{>{\raggedright}p{#1}}
\newcolumntype{C}[1]{>{\centering}p{#1}}
\newcolumntype{R}[1]{>{\raggedleft}p{#1}}
\newcolumntype{D}{>{\centering\arraybackslash}X}

\definecolor{darkgreen}{rgb}{0,0.5,0}

\newcommand{\edit}[1]{{\color{black}{#1}}}

\begin{document}
\title{The surface code with a twist}
\author{Theodore J. Yoder}
\thanks{tjyoder@mit.edu}
\affiliation{Department of Physics, Massachusetts Institute of Technology}
\author{Isaac H. Kim}
\affiliation{IBM, Thomas J. Watson Research Center}

\begin{abstract}
The surface code is one of the most successful approaches to topological quantum error-correction. It boasts the smallest known syndrome extraction circuits and correspondingly largest thresholds. Defect-based logical encodings of a new variety called twists have made it possible to implement the full Clifford group without state distillation. Here we investigate a patch-based encoding involving a modified twist. In our modified formulation, the resulting codes, called \emph{triangle codes} for the shape of their planar layout, have only weight-four checks and relatively simple syndrome extraction circuits that maintain a high, near surface-code-level threshold. They also use $25\%$ fewer physical qubits per logical qubit than the surface code. Moreover, benefiting from the twist, we can implement all Clifford gates by lattice surgery without the need for state distillation. By a surgical transformation to the surface code, we also develop a scheme of doing all Clifford gates on surface code patches in an atypical planar layout, though with lower qubit efficiency than the triangle code. Finally, we remark that logical qubits encoded in triangle codes are naturally amenable to logical tomography, and the smallest triangle code can demonstrate high-pseudothreshold fault-tolerance to depolarizing noise using just 13 physical qubits.
\end{abstract}

\maketitle

\section{Introduction}
The surface code \cite{Bravyi1998,Dennis2002,Fowler2012c} is a dominating proposal for nearest-neighbor quantum error-correction in a plane. But there are some good reasons for its ubiquity. For instance, asymptotically in the code distance and up to a constant factor, the surface code uses the fewest number of qubits per logical qubit. Also, the surface code has optimally sized checks (weight-four) for a topological stabilizer code \cite{Aharonov2011} and a simple scheme for syndrome extraction using only one ancilla qubit per check \cite{Tomita2014}. Moreover, the syndrome information can be processed efficiently \cite{Dennis2002,Wang2011,Hutter2014,Bravyi2014}, resulting in the highest known topological fault-tolerance threshold \cite{Wootton2012}. The huge body of work on the surface code inspires optimism -- perhaps we have found the ``best" topological code \cite{Fowler2012a}.

Of course, this is a difficult claim to justify completely in any rigorous sense. There are a plethora of other topological coding strategies offering advantages and disadvantages. As one example, subsystem topological codes can perform syndrome measurement by measuring only operators that are less than weight four \cite{Bombin2010a,Bravyi2012}. As another example, color codes \cite{Bombin2006} can implement Clifford gates transversally \cite{Bombin2015,Kubica2015a} and thus more efficiently than the surface code, even if they fail to achieve as large a threshold \cite{Landahl2014}. Strictly speaking there is not even just one best strategy for computing with the surface code. One option is to encode multiple qubits into one large region of surface code by using defects called holes \cite{Raussendorf2007}. Braiding the holes results in logical operations on the encoded qubits \cite{Bombin2009,Fowler2012c}, and, using the technique of magic-state distillation \cite{Bravyi2005,Fowler2013}, can achieve universality \cite{Fowler2009} limited only by $T$-depth of the circuit \cite{Fowler2012b}. Alternatively, computing with surface code ``patches'', where each logical qubit is localized to a square grid of physical qubits, can be done with lattice surgery \cite{Horsman2012,Landahl2014} and comparatively few qubits. Still, both these strategies use state distillation for the Clifford phase gate, $S=\text{diag}(1,i)$.

Surface code computation has more recently undergone another reformulation with the introduction of a different type of defect, called a \emph{twist} \cite{Bombin2010b,Hastings2014}. A twist defect is a weight-five check embedded in the surface code lattice. Unlike a hole, a twist destroys the Calderbank-Shor-Steane (CSS) \cite{Calderbank1996,Steane1996} nature of the surface code, as the weight five check contains Pauli $Y$. Four (or three) twists in a surface code lattice with uniform boundary can encode one qubit \cite{Hastings2014}. Interestingly, this twist encoding appears to have advantages over the hole encoding, as the full Clifford group can be fault-tolerantly implemented by local operations and without state distillation for the phase gate $S$ \cite{Hastings2014}. Using both the hole and twist encodings together is also possible \cite{Brown2016} and implements the same gate set similarly. So far, however, while the advantages of twist defects have been explored for multiple defects within the same large lattice, there has been little said about possible advantages for patch-based schemes of computation, which, due to lower qubit overhead, are more friendly to current experiments.

This is our focus in this work, the development of a planar patch layout that uses twist defects to achieve full Clifford group computation with local operations. Universality can then be achieved by injecting and distilling magic states for $T=\sqrt{S}$ gates. We find that a single twist defect placed in the middle of a patch of surface code suffices for this task. Moreover, the typical formulation of a twist as a weight-five check can be simplified in our case to weight-four. The family of non-CSS codes corresponding to these twist-containing patches we call \emph{triangle codes}, because the physical qubits making up a single patch can be arranged to fit within a triangle in the plane. We also show how triangle codes are related by lattice surgery to the traditional surface code, and use this to deduce a surgical method for implementing $S$ on patches of surface code in a nonstandard planar layout.

In addition to providing a full, distillation-free implementation of the Clifford group, triangle codes have further advantages over patches of surface code. First, triangle codes of (odd) distance $d$ use $3d^2/4+1/4$ data qubits per logical qubit, beating the (rotated \cite{Bombin2007,Tomita2014}) surface code's $d^2$. While this does not asymptotically beat some color codes \cite{Landahl2014}, which are yet another constant factor better, triangle codes also benefit from near surface code thresholds. Indeed, depolarizing-noise fault-tolerant circuits for syndrome extraction on triangle codes can be made nearly as simple as that on surface codes, using the minimal number of timesteps when amortized over many rounds of extraction and nearly the minimal number of ancillas (i.e.~one per stabilizer generator).

A second advantage of triangle codes is their symmetry -- logical Pauli operators $\bar X$, $\bar Y$, and $\bar Z$ can be taken to each lie along a different side of the triangular layout of qubits. This means that fault-tolerant measurement and initialization in any Pauli basis is possible, albeit not as straightforwardly as on CCS codes like the surface code. Additionally, triangle codes possess transversal order-3 single-qubit Clifford gates, which cyclically permute the Paulis, up to a permutation. For a single patch of triangle code, the fact that a permutation is required can be ignored, and a logical order-3 gate, plus logical initialization and measurement of the patch, suffice for tomography (or randomized benchmarking) of the logical qubit. Noting that for $d=3$ this can be done with just 13 physical qubits including ancillas makes this a promising scheme for early implementations of complete quantum fault-tolerance to depolarizing noise.

In Sec.~\ref{sec:construction} we define the triangle codes, calculate their threshold for topological memory, and present circuits for syndrome extraction. In Sec.~\ref{sec:init_meas}, we discuss initialization and measurement of triangle code patches. The strategies we put forward there translate into schemes for computation with tesselating triangle patches in the plane, Sec.~\ref{sec:planar_comp}. In Sec.~\ref{sec:distance_3} we discuss circuits specifically for the smallest distance-3 triangle code, and calculate pseudothresholds. Sec.~\ref{sec:conclusion} concludes.

\section{Constructing and error-correcting triangle codes}\label{sec:construction}
\begin{figure}
\includegraphics[width=\columnwidth]{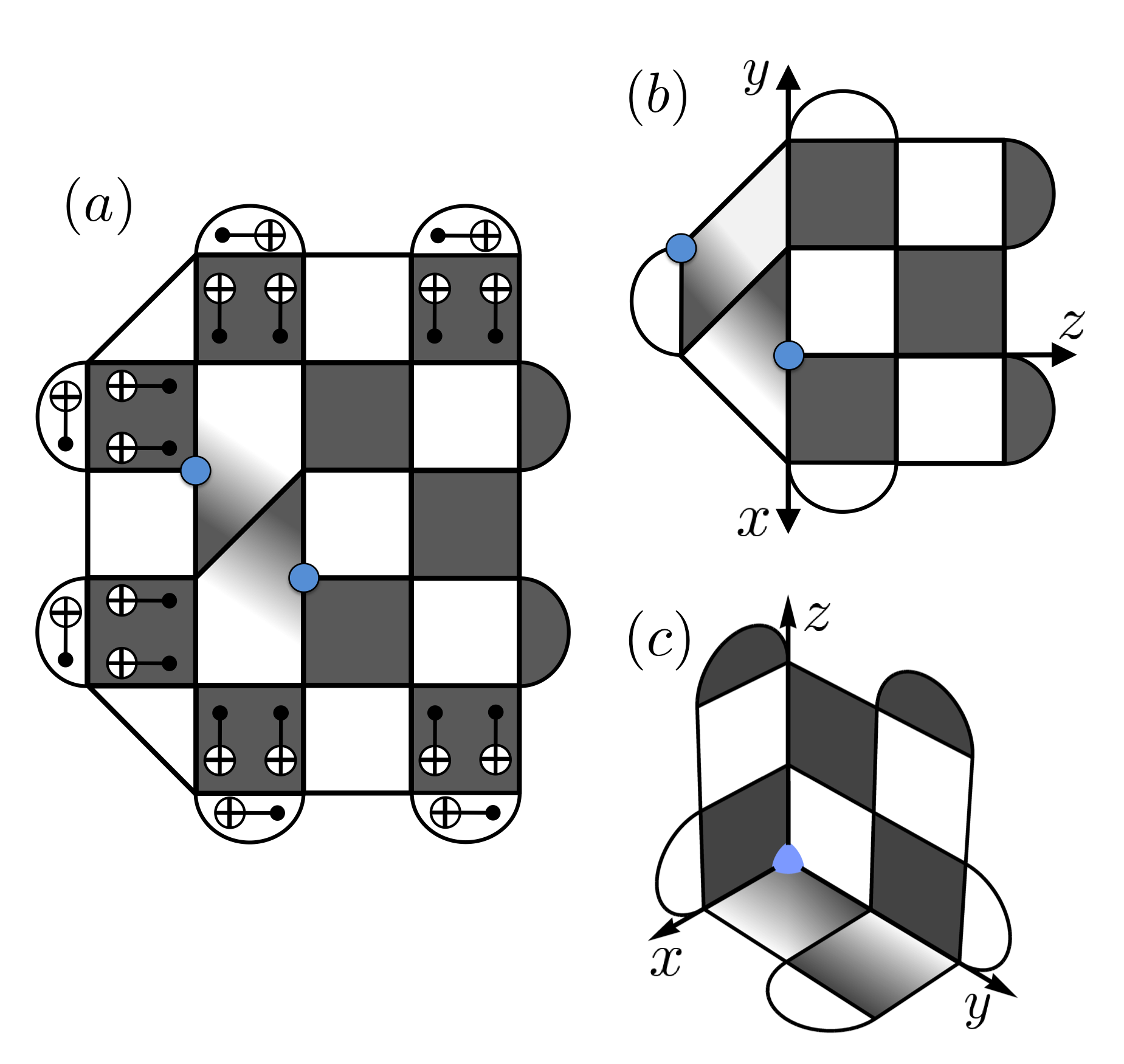}
\caption{\label{lattsurg_disloctotriangle} Converting a dislocation code \cite{Bombin2010b,Hastings2014} into an asymmetric triangle code. In (a) a dislocation code with two twists encodes one logical qubit (physical qubits are on the vertices of the lattice pictured). Stabilizers, both plaquettes and loops, are colored dark gray if they are $X$-type and white if they are $Z$-type. Stabilizers graded white to gray are mixed-type, and consist of $X$s where they are gray, $Z$s where they are white, and a $Y$ on the blue, dotted qubits. Overlaid is a local Clifford circuit of CNOTs that takes the code to (b) when the inner CNOTs are applied, followed by the outer. Note that many qubits are unentangled from the bulk of the code during this process. From (b) to (c), a single-qubit Clifford is applied to remove a Pauli $Y$ from one stabilizer, and the code is reoriented to get an asymmetric triangle code.}
\end{figure}

Our first task is to construct the family of triangle codes. The signature element of a triangle code is a central twist defect, and so to begin, we show in Fig.~\ref{lattsurg_disloctotriangle} how the triangle codes arise from the dislocation codes \cite{Bombin2010b,Hastings2014} by lattice surgery. Notice that this surgery simplifies the code --- all stabilizer generators are now at most weight-4 and fewer physical qubits are required. Further lattice surgery can symmetrize or extend the triangle code into the more general family that we describe now.

\begin{figure}
\includegraphics[width=\columnwidth]{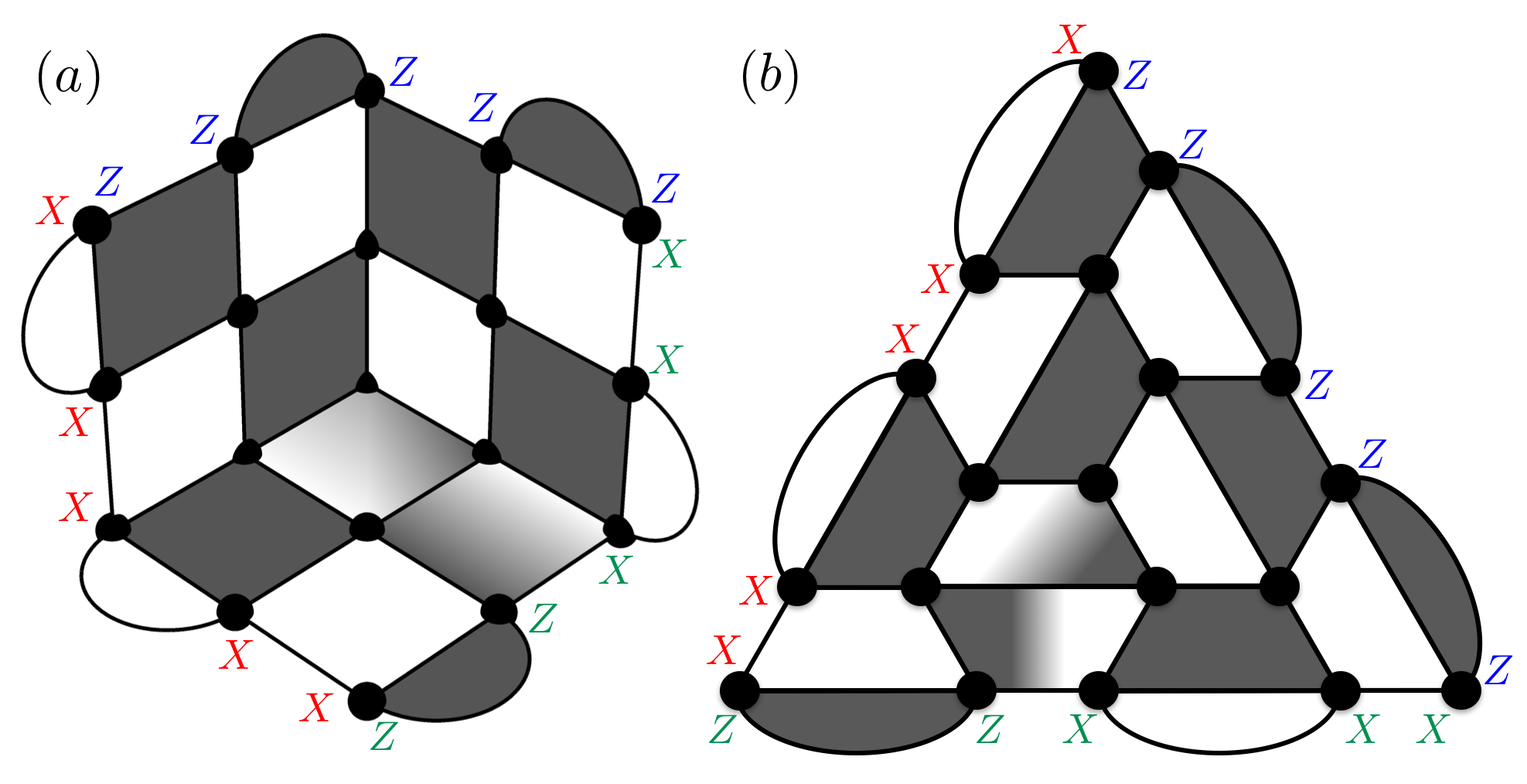}
\caption{\label{intro_triangleplanar} (a) The distance-5 triangle code viewed as three conjoined surface codes in 3D. Black dots are data qubits, and stabilizers are colored as in Fig.~\ref{lattsurg_disloctotriangle} except for the Pauli $Y$ on the origin that is part of the central mixed-type stabilizer, which is now left implicit. Along the sides, a string of $Z$s (blue) is $\bar Z$, a string of $X$s (red) $\bar X$, and a mixed-type string (green) $\bar Y$. (b) Shows the same, but in a plane.}
\end{figure}

A general $r\times s\times t$ triangle code is described by three positive integer parameters $r,s,t$. We find the codes are most easily described by placing qubits at points in three dimensions, $(x,y,z)\in\mathbb{Z}_r\times\mathbb{Z}_s\times\mathbb{Z}_t$ for $\mathbb{Z}_n=\{0,1,\dots,n-1\}$, subject to the constraint that at least one of $x,y,z$ is zero. In other words, this places qubits on integer lattices in the $xy$-, $yz$-, and $xz$-planes as seen in Fig.~\ref{intro_triangleplanar}(a). Note that placing qubits in three dimensions is entirely for conceptual simplicity, and a projection into two dimensions is readily obtained, Fig.~\ref{intro_triangleplanar}(b). The projection fits within the eponymous triangle.

Stabilizers of a triangle code are local Pauli operators of either weight-4 (plaquettes) or weight-2 (loops). Plaquettes are associated with half-integer lattice points in the $xy$-, $yz$-, and $xz$-planes. They may be $X$-type (consisting of only Pauli $X$ and $I$ operators), $Z$-type (only Pauli $Z$ and $I$), or mixed-type (any other combination of Paulis). By convention, we will choose the plaquette at $(1/2,0,1/2)$ to be $X$-type. Plaquettes, in order to commute, must alternate $X$-type and $Z$-type whenever possible (e.g.~implying the plaquette at $(0,1/2,1/2)$ is $Z$-type), but geometry demands that mixed-type plaquettes are placed starting from the origin and extending in some direction to the outer edge. We conventionally take the plaquettes associated to the points $(1/2,m+1/2,0)$, for $m\in\mathbb{Z}_{s-1}$ to be mixed type. The central mixed-type plaquette at $(1/2,1/2,0)$ is special in that it contains Pauli $Y$ on the qubit $(0,0,0)$, while all other mixed-type plaquettes are half $X$ and half $Z$. 

Loops are placed with support on every other pair of qubits along the boundary. They can also be $X$-type, $Z$-type, or mixed-type. By convention, we choose $Z$-type loops to be associated with qubits for which $x=r-1$. This fixes the position and type of all other loops.

By counting qubits and stabilizer generators it can be checked that triangle codes for any $r,s,t$, encode one logical qubit. Logical operators $\bar X$, $\bar Y$, and $\bar Z$ may be taken to lie on the boundary, crossing the $x$-, $y$-, and $z$-axes, respectively, as shown in Fig.~\ref{intro_triangleplanar}(a). If $P_{(x,y,z)}$ denotes a Pauli acting on the qubit at $(x,y,z)$,
\begin{align}
\bar X&=\prod_{j\in\mathbb{Z}_s\setminus0}\hspace{-4pt}X_{(r-1,j,0)}\prod_{k\in\mathbb{Z}_t}\hspace{-2pt}X_{(r-1,0,k)},\\
\bar Y&=\prod_{i\in\mathbb{Z}_r\setminus0}\hspace{-4pt}Z_{(i,s-1,0)}\prod_{k\in\mathbb{Z}_t}\hspace{-2pt}X_{(0,s-1,k)},\\
\bar Z&=\prod_{i\in\mathbb{Z}_r\setminus0}\hspace{-4pt}Z_{(i,0,t-1)}\prod_{j\in\mathbb{Z}_s}\hspace{-2pt}Z_{(0,j,t-1)}.
\end{align}
The minimum weights of the logical operators are respectively, $d_x=s+t-1$, $d_y=r+t-1$, and $d_z=r+s-1$. We describe symmetric triangle codes that have parameters $r=s=t=(d+1)/2$ by the adjective distance-$d$.

\begin{figure}
\includegraphics[width=\columnwidth]{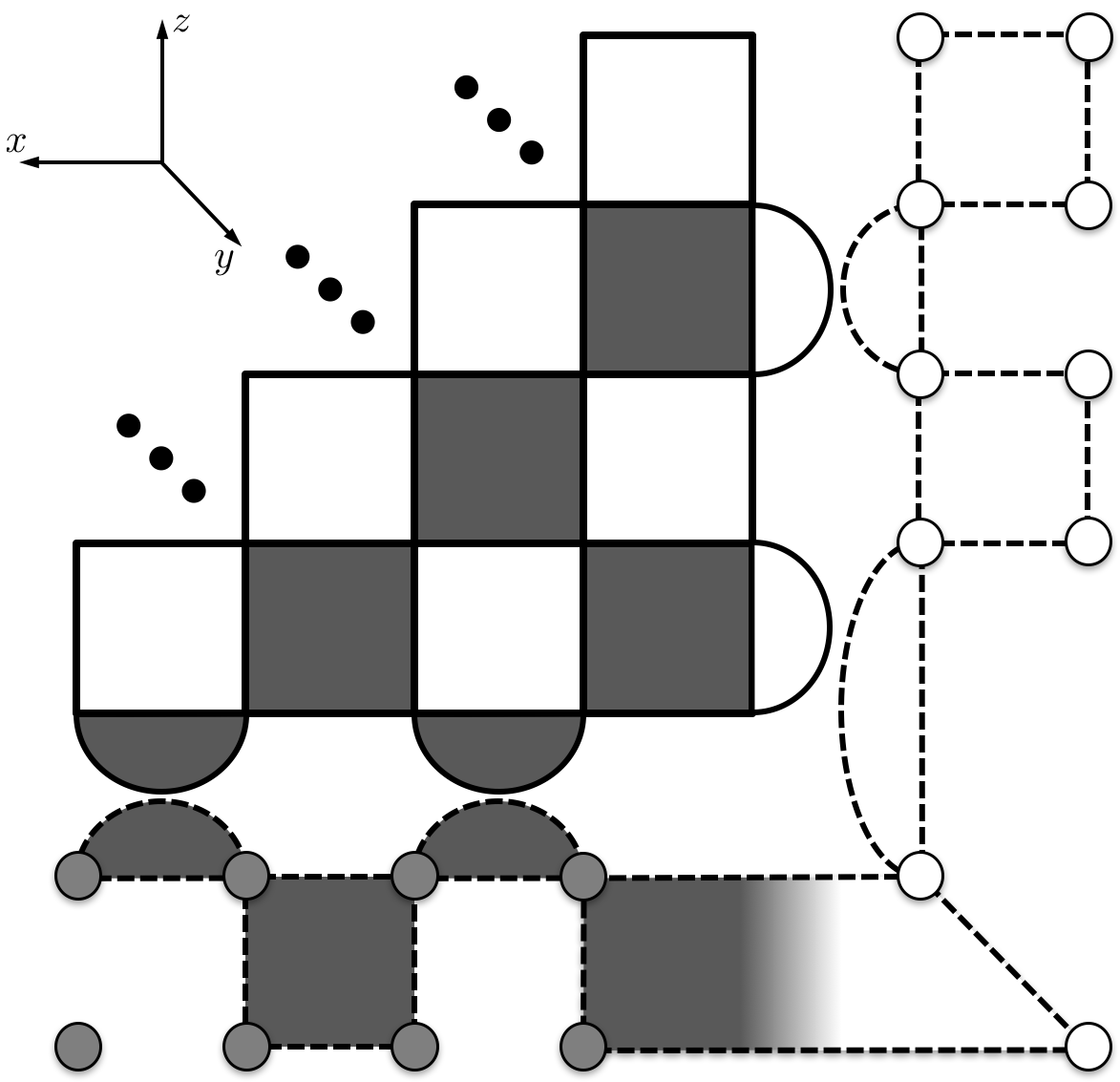}
\caption{\label{lattsurg_cornercreation} Creating a triangle code from a $d\times d$ surface code by extending from a corner. In the diagram, ancilla qubits are prepared in $\ket{+}$ (gray) or $\ket{0}$ (white). Then, the stabilizers of the triangle code are measured $O(d)$ times. The code is always in the $+1$-eigenstate of the dashed plaquettes and the product of dashed loops and solid loops that together make up a plaquette of the triangle code. Thus, syndromes from half of the triangle code stabilizers in the $xy$- and $yz$-planes can be used reliably for decoding. The effective distance of this procedure is $d$.}
\end{figure}

Triangle codes may also be more directly constructed from a surface code (and vice versa). This is not entirely surprising when the corners of the surface code are viewed as twists \cite{Brown2016}. A fault-tolerant method to perform this conversion is shown in Fig.~\ref{lattsurg_cornercreation}. Indeed, in our notation and with our conventions, a $d\times d$ surface code is precisely a $d\times1\times d$ triangle code.


\begin{table}
\centering
\begin{tabular}{|c||c|c|c|}
\hline
 code & syndrome & X threshold & X-Z threshold \\ \hline\hline
 triangle & ideal & $\approx 10\%$ & $\approx 10\%$ \\
 rotated & ideal & $\approx 10\%$ & $\approx 10\%$ \\ \hline
 triangle & noisy & $\approx 3.2\%$ & $\mathbf{\approx 2.6\%}$ \\
 rotated & noisy & $\approx 3.2\%$ & $\mathbf{\approx 3.2\%}$ \\
 \hline
\end{tabular}
\caption{Numerical threshold estimates for the triangle and rotated surface codes show appreciable differences only for noisy syndrome measurements and bit-phase flip noise (bold).\label{tab:phenom}}
\end{table}

So far we have described triangle codes as topological codes, but we have not yet proved that a threshold for storing quantum information exists. In Appendix~\ref{app_thresh}, we show such a threshold exists through a straightforward argument analogous to that for the toric code \cite{Dennis2002}. For the purpose of comparison with the surface code, we also estimate thresholds for the triangle code and rotated surface code \cite{Tomita2014} families under phenomenological noise. We consider both bit (X) and bit-phase (X-Z) flip storage errors and measurement models where the measured syndromes are ideal or noisy. Table~\ref{tab:phenom} summarizes the results, and Appendix~\ref{app_thresh} provides more detail on the simulation.

We now turn to syndrome extraction circuits for the triangle codes. The key notion here is that of effective distance. We say that the effective distance of a syndrome extraction circuit is $d$ if no fewer than $d$ faulty circuit components (e.g.~single-qubit gates, two-qubit gates, preparations, measurements) can create a logical error on the data, \edit{while at the same time a trivial syndrome is measured}. This implies being able to correct any set of $\lfloor d/2\rfloor$ faults given the syndrome, because no two such sets can have the same syndrome. We will assume the depolarizing noise model for faults when designing these circuits.

A $d\times d$ surface code possesses a very simple method \cite{Tomita2014} for syndrome extraction with effective distance $d$ under depolarizing noise, using an optimal number of ancilla qubits -- one per stabilizer check -- and an optimal number of timesteps -- six, if we say coupling the ancilla to all four data qubits takes four total timesteps and initialization and measurement each take another. This procedure can leave two-qubit errors on the data from a single fault (commonly called a ``hook" error \cite{Dennis2002}) but such errors are oriented such that $d$ hook errors are required to write a logical error onto the data (see Appendix~\ref{app_synextract}).

We might hope that the triangle code, being for the most part surface code, also supports as simple a syndrome extraction. This is nearly the case. By brute-force check of the distance-3 symmetric triangle code, we actually find no fault-tolerant procedure using both the minimal number of qubits and the minimal number of timesteps. This suggests we have to think a little more creatively to find a simple syndrome extraction circuit for the triangle codes. Luckily, we do not have to use too many more resources. Two qubits in addition to the allotted one per stabilizer check suffice to perform full-distance extraction on the triangle code. In fact, amortized over many rounds of syndrome extraction, our protocol also uses the minimal number of timesteps per round, six.

\begin{figure}
\includegraphics[width=\columnwidth]{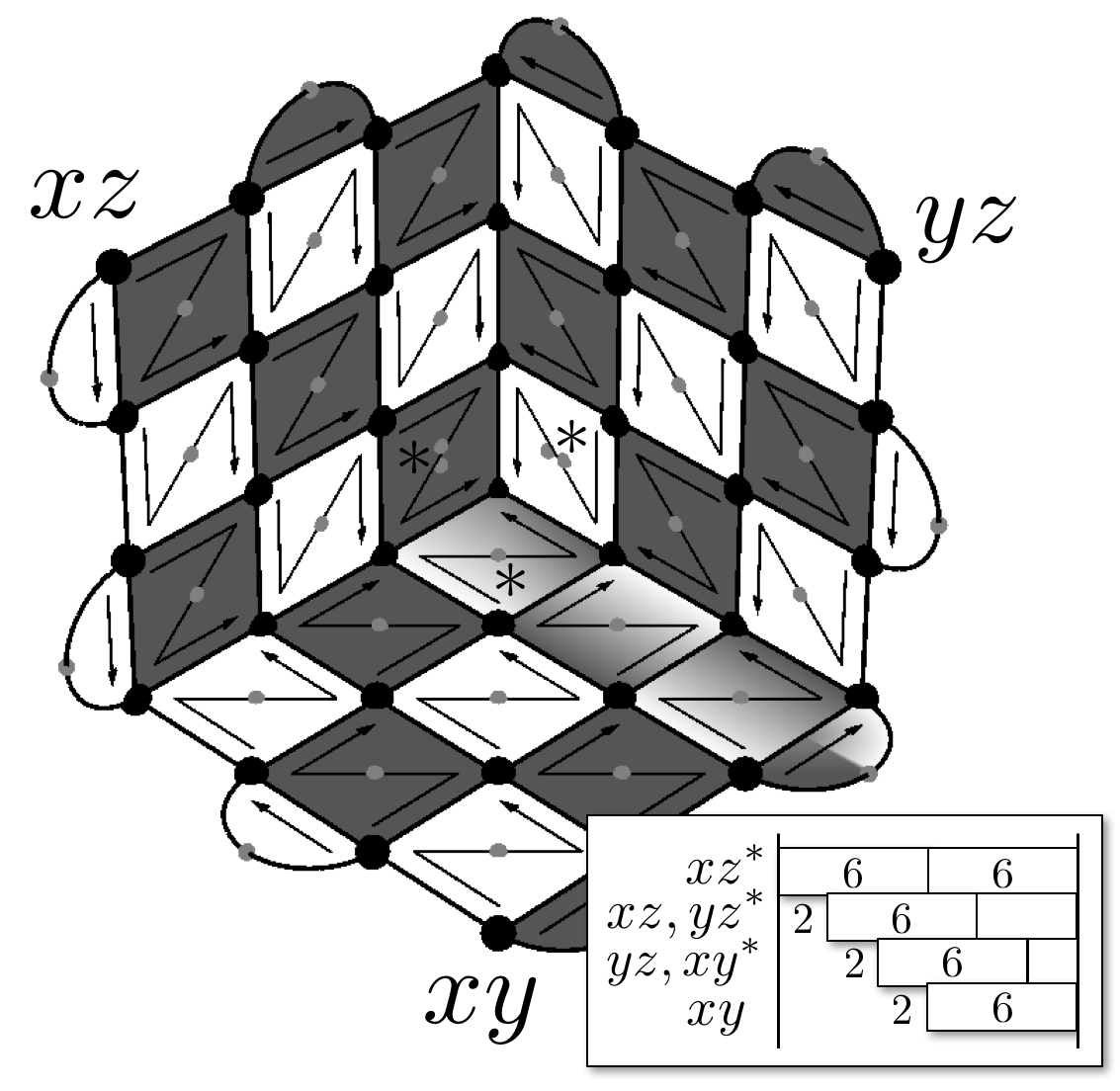}
\caption{\label{circ_synextract} A schematic of the syndrome measurement on the triangle code. The arrows show the order (tail to head) in which an ancilla couples to the data after it is initialized. Two plaquettes adjacent to the pivot are endowed with flag qubits to detect hook errors, as described in Fig.~\ref{circ_flags}. The inset shows that syndrome extraction can be fit together so, over many cycles, the amortized time per complete syndrome extraction is six timesteps. The three planes (labeled $xz$, $yz$, $xy$) are measured two timesteps offset and starred (*) stabilizers, the plaquettes adjacent to the origin, begin their measurement cycle two timesteps earlier than the rest of their plane to avoid time conflicts when coupling to the data.}
\end{figure}

Our syndrome extraction circuit is shown in Fig.~\ref{circ_synextract}. The design uses the standard scheduling of the surface code \cite{Tomita2014} within a plane, and staggers the timing of different planes to avoid conflicts in coupling to the data qubits. To achieve fault-tolerance to the maximum number of depolarizing faults, even this is not enough, however. For example, just two hooks can cause the weight three error $Z_{(1,0,0)}Z_{(0,0,0)}Z_{(0,1,0)}$ and the remaining $d-3$ Paulis making up $\bar Z$ (considered equivalently with support across the $x$- and $y$-axes) can be caused by single qubit faults. Thus, $\bar Z$ may be written onto the data with only $d-1$ faulty circuit components. On the other hand, reducing the distance by one is actually the worst that can happen in this design (see Appendix~\ref{app_synextract}). A larger triangle code using $3(d+1)^2/4+1/4$ qubits (for even $d$), still with asymptotic qubit count of $3d^2/4$, will perform at effective distance $d$ even when syndromes are extracted using just one ancilla per check.

\begin{figure}
\includegraphics[width=\columnwidth]{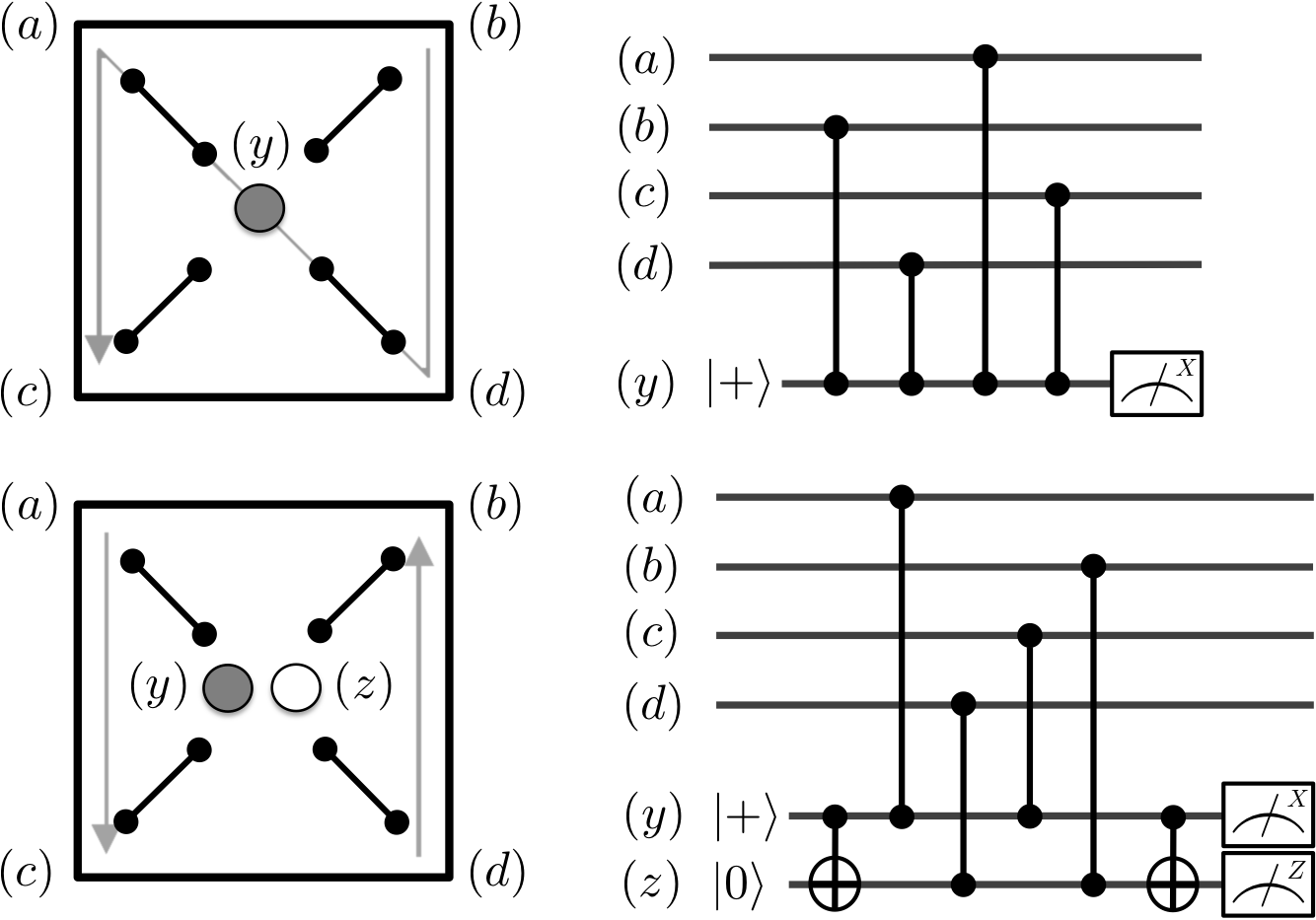}
\caption{\label{circ_flags} The conventional single-qubit syndrome measurement (top row) and the flagged syndrome measurement (bottom row) both take six timesteps (including preparation and measurement). The gates $\bullet\text{---}\bullet$ in this example of measuring a $Z$-type plaquette are controlled-$Z$. In Fig.~\ref{circ_synextract}, we suggest using the flag qubits on two plaquettes near the pivot to achieve full-distance syndrome extraction.}
\end{figure}

If we wish to make syndrome extraction for a distance-$d$ triangle code with effective distance $d$, we can detect hook errors by creating and decoding 2-CATs (see Fig.~\ref{circ_flags}) for (any) two of the plaquettes adjacent to the origin. This efficiently deals with the case noted above, where a pair of hook errors near the origin can write three errors onto the data, and is represented in Fig.~\ref{circ_synextract} as a pair of ancillas in two of the central plaquettes.

Neither syndrome extraction with 2-CATs nor extraction on a larger code are ideal for distance three, if the goal is to make the most qubit-efficient fault-tolerant architecture with high pseudothreshold. In Sec~\ref{sec:distance_3}, we create a syndrome extraction circuit specialized to distance three using one ancilla per check.

\section{Initialization and measurement}\label{sec:init_meas}
The fact that triangle codes are not CSS complicates their initialization and measurement protocols. This is actually the biggest challenge to computing with them. Gates, in contrast, which are discussed in detail in Sec.~\ref{sec:planar_comp}, can be performed with a combination of known lattice surgery techniques \cite{Horsman2012} supplemented with 1-bit teleportation \cite{Gottesman1999} and, for non-Cliffords, magic-state injection and distillation \cite{Bravyi2005}.

As an example of the complications that arise in measuring a non-CSS code, consider attempting to deduce the eigenvalue of $\bar Z$ by measuring all qubits of a symmetric, distance-$d$ triangle code in the $Z$-basis. The parity of the qubits across the top of the code should equal $\bar Z$. But can we error-correct this value with maximum distance? In fact, we cannot. Because we measured all qubits in the $Z$-basis, we are lacking the information from the central mixed-type stabilizer. Thus, a string of $X$ errors from the origin to the top of the code is an undetectable weight $(d+1)/2<d$ error that causes our measurement to fail. \edit{In general, if we have any syndrome measurement scheme that fails to report all the values in a complete set of stabilizers, we can find a string of errors from the unreported stabilizer to an edge (sometimes using the central twist or dislocation to change type) that has weight less than $d$ and anticommutes with the logical operator being measured. As a special case, performing logical measurement by measuring individual qubits only will certainly leave a stabilizer unreported and so cannot achieve maximum distance.} Thus, we can either be satisfied with a noisy, but simple, logical measurement (perhaps in an implementation with extremely reliable single-qubit measurements), or develop full-distance alternatives.

We present three different approaches that achieve full-distance initialization and measurement. Each approach results in a different scheme in Sec.~\ref{sec:planar_comp} for computing with the triangle codes. We feel each has something to offer and we discuss tradeoffs later. \edit{The first two are motivated by the correspondence of the center of the triangle code with a twist (see Sec.~\ref{sec:construction}), and the association of corners of the surface code with twists \cite{Brown2016}. The last is a simple use of Shor's measurement \cite{Shor1996} with locality constraints.} In Sec.~\ref{sec:distance_3}, we focus on the distance-3 triangle code, for which we have developed specialized alternatives to these schemes.

\edit{Let us briefly describe and name each approach to initialization and measurement.} The first we call code conversion \edit{(CC)}, because it works by using Fig.~\ref{lattsurg_cornercreation} to convert between the surface code, for which initialization and measurement procedures are simple and known, and the triangle code. This actually moves any logical state between the surface code and the triangle code. Notice, however, that this is stronger than necessary if all we want is to initialize or measure a Pauli eigenstate. For this reason, the second method, called basis-state conversion \edit{(BC)}, uses fewer qubits to convert a Pauli basis state from the surface to triangle codes or measure the basis-state transversally. The final method, employing CAT states \edit{(CS)}, is the most qubit efficient but least time efficient. \edit{We detail each of these in turn now.}

\edit{To understand the CC approach}, notice that a distance $2d+1$ triangle code admits transversal measurement of $\bar X$, $\bar Y$, and $\bar Z$ with effective distance $d$ by simply measuring all qubits in the $X$-, $Y$-, or $Z$-bases. Initialization of a state in the distance $2d+1$ triangle code can also be done by initializing a $d\times d$ surface code patch (for which there are known procedures \cite{Fowler2012a,Tomita2014} with effective distance $d$) and then extending the patch using Fig.~\ref{lattsurg_cornercreation}. This procedure is very much based on the surface code, and so storing data with distance $\ge d$ could be done without ever converting to the triangle code at all. However, the embedding of the surface code patch in a large triangular patch is important for performing single-qubit Clifford gates, as explained in Section~\ref{sec:planar_comp}.

\begin{figure}
\hspace*{0.1in}
\includegraphics[width=0.85\columnwidth]{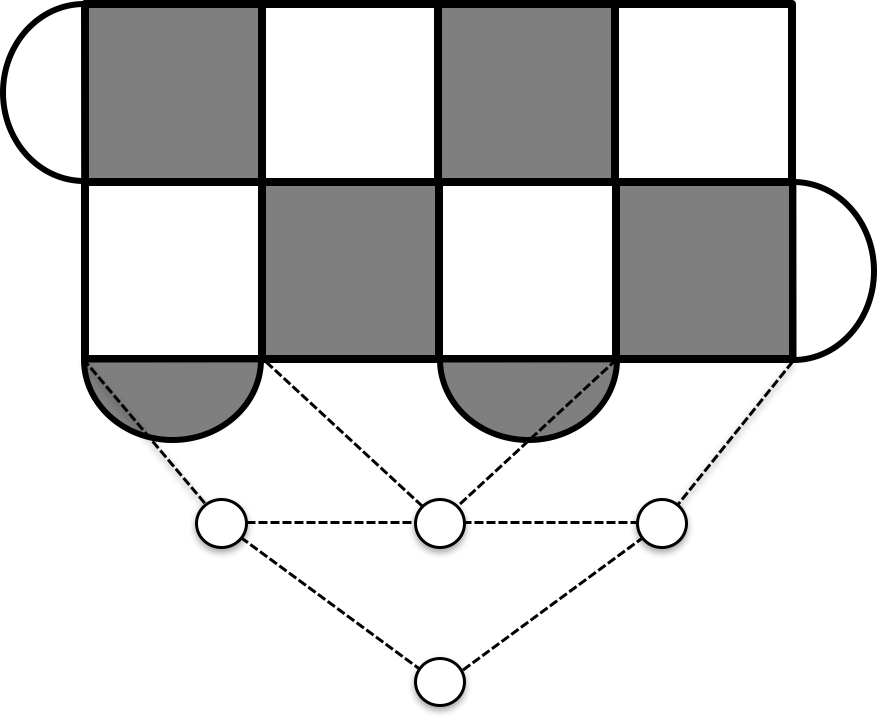}
\caption{\label{lattsurg_basisstateconv} Converting a $d\times d$ surface code (in this case $d=5$ with the top two rows of qubits omitted) to a $(d+1)/2\times(d+1)/2\times d$ triangle code. To do so, stop measuring the bottom loops (here the two $X$-type ones) and start measuring the stabilizers of the triangle code (dashed skeleton) for $O(d)$ cycles. The initial state of the new qubits is not relevant to the effective distance, but one might be able to correct more higher weight errors if they were initialized as, for instance, a collection of 2-qubit CAT states or as a $(d-1)/2\times(d-1)/2$ surface code.}
\end{figure}

\edit{In our next approach, BC}, we also prepare a basis state, say $\ket{\bar0}$, in a $d\times d$ surface code, but instead of treating this surface code as $1/3$ of a triangular patch, we treat it as $2/3$, the $xz$- and $yz$-planes say. Conversion to the surface code can then be done by ceasing measurement of the loop operators along the bottom of the surface code, and measuring the stabilizers of a $(d+1)/2\times(d+1)/2\times d$ triangle code $O(d)$ times (see Fig.~\ref{lattsurg_basisstateconv}). Regardless of the initial state of the new qubits, this process has effective distance $d$, since $\bar X$ must always cross the original $d\times d$ surface code patch, whose $Z$-type stabilizers are always reliable throughout the conversion. A $\bar Z$ error is ignorable since we are preparing $\ket{\bar0}$. After this conversion, symmetrize the triangle code, so that we are left with the symmetric, distance-$d$ code, by measuring its stabilizers and ceasing to measure the extraneous stabilizers in the topmost rows. Again, the logical operator $\bar Z$ can occur from fewer than $d$ faults but is irrelevant. Logical measurement is done by inverting the last step --- extend the triangle code along the $z$-axis $d/2$ steps, and measure all qubits in the $Z$-basis.

\edit{The CS} method \cite{Shor1996} to initialization and measurement is probably the most straightforward. Using a row of $d$ ancilla qubits along an edge of a distance-$d$ triangle code, we can create and verify a $d$-CAT state for measuring the logical operator also lying along that edge. Creating the $d$-CAT with local operations takes $d$ timesteps (e.g.~by measuring two-bit parity checks starting from $\ket{+}^{\otimes d}$ \cite{Brooks2013}), and during this process, we must be collecting syndromes from the patch. We must also repeat the CAT state measurement $O(d)$ times to reliably measure the logical state, leading to $O(d^2)$ time for logical measurement. Initialization is essentially the same process, using $O(d^2)$ time to measure a logical operator $O(d)$ times and the stabilizers $O(d^2)$ times to project the code onto the desired logical state.

\section{Planar, nearest-neighbor computation with triangles}\label{sec:planar_comp}
\begin{figure}
\includegraphics[width=\columnwidth]{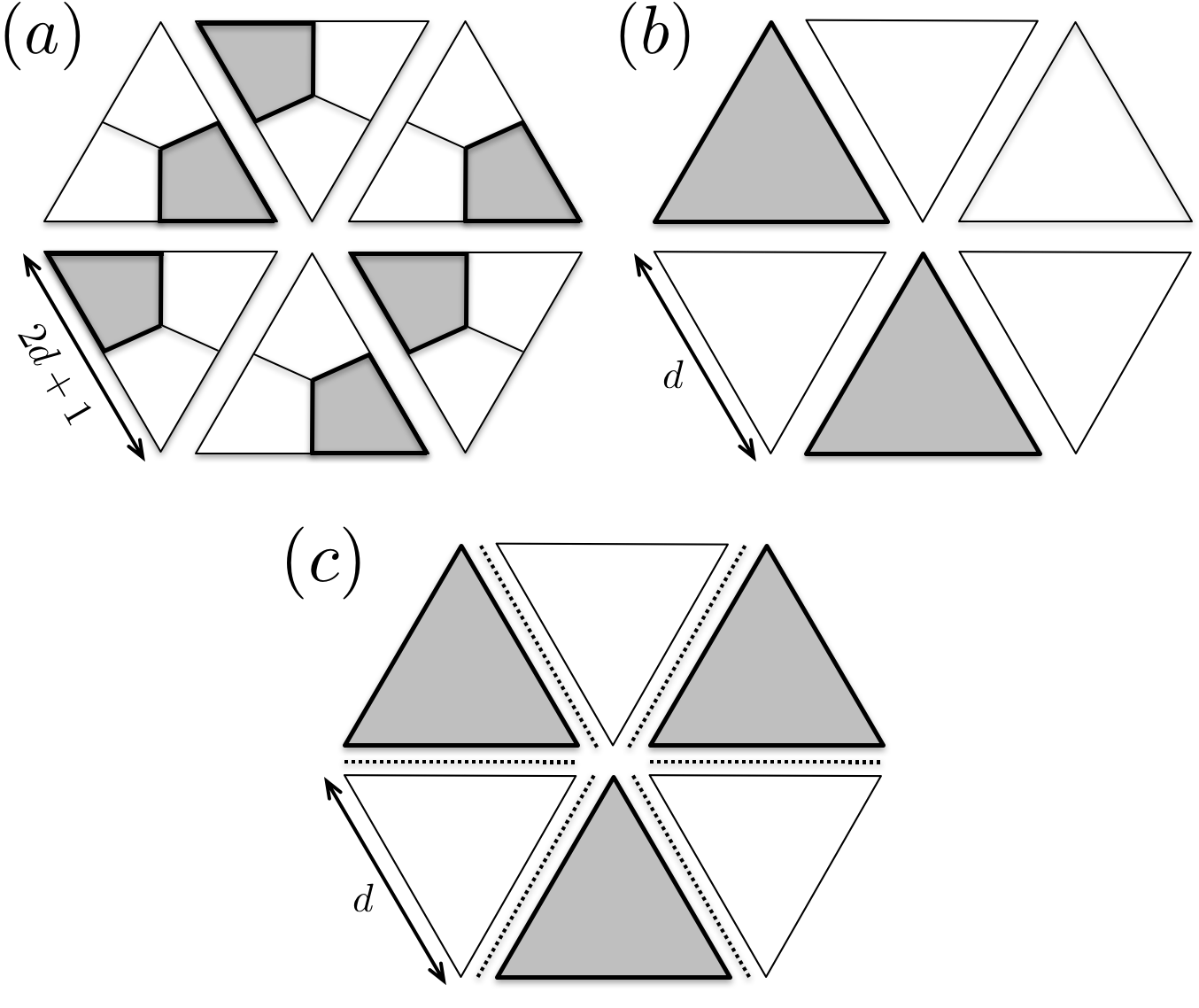}
\caption{\label{lattsurg_3layouts} Tessellating patches of triangle code capable of complete Clifford computation with effective distance $d$, corresponding to using (a) code conversion \edit{(CC)} (b) basis-state conversion \edit{(BC)} and (c) CAT states \edit{(CS)} for initialization and measurement. In a resting configuration, the shaded regions represent patches of physical qubits that are encoding data, while blank areas represent ancillary patches. In (a) the resting code is actually the surface code, and the triangular patches (with qubit layout as in Fig.~\ref{intro_triangleplanar}(b)) are present only to facilitate single-qubit gates. In (b), we use two ancilla patches per data patch to perform gates, because initialization and measurement use $2/3$ of a neighboring patch. In (c), we add ancillary qubits between patches to make CAT states for initialization and measurement, and so require just one ancillary patch per data patch for arbitrary reorientation and CNOTs with neighboring data patches.}
\end{figure}

With initialization and measurement protocols created, our next priority is to perform the Clifford group on triangle codes. Inevitably, by desiring to implement the two-qubit gate CNOT, we are forced to consider the layout of several triangle codes within the same planar geometry. The basic design is to tile the plane with equilateral triangles, each having the potential to encode a qubit as per Fig.~\ref{intro_triangleplanar}(b). However, the size of these code patches, and also which patches encode computational data and which must act as ancillas used for gates, varies depending on the initialization and measurement protocol used from Sec.~\ref{sec:init_meas}. So, we present three designs that vary in these ways, and thus also vary in how space efficiently data is stored and how time efficiently gates are performed on that data (see Table~\ref{tab:compare_layouts} and Fig.~\ref{lattsurg_3layouts}).

\setlength\extrarowheight{3pt}
\begin{table}
\setlength{\tabcolsep}{0.4em}
\begin{tabular}{|c||c|c|c|}
\hline
init. \& meas. scheme & qubits/log. & $H,S$ & CNOT \\
\hline\hline
code conversion \edit{(CC)} & $3d^2+O(d)$ & $O(d)$ & $O(d)$ \\
basis-state conv.~\edit{(BC)} & $9d^2/4+O(d)$ & $O(1)$ & $O(d)$ \\
CAT states \edit{(CS)} & $6d^2/4+O(d)$ & $O(1)$ & $O(d^2)$ \\
\hline
\end{tabular}
\caption{\label{tab:compare_layouts} Comparing three strategies for planar Clifford computation in triangle code geometry with distance $d$. Compared quantities are qubit counts per logical qubit (including ancillas needed for gates) and Clifford gate times.}
\end{table}

In Fig.~\ref{lattsurg_3layouts}(a), we show the first of our three designs corresponding to the \edit{CC method of initialization and measurement}. Indeed, in this design we actually prefer to use the surface code ($1/3$ of a triangular patch) as our resting code, holding the data between gates. Standard lattice surgery \cite{Horsman2012} suffices for performing CNOT gates between the surface code patches. However, the large triangular patches become important when we wish to perform single-qubit Clifford gates.

\begin{figure}
\hspace*{0.04in}
\includegraphics[width=0.95\columnwidth]{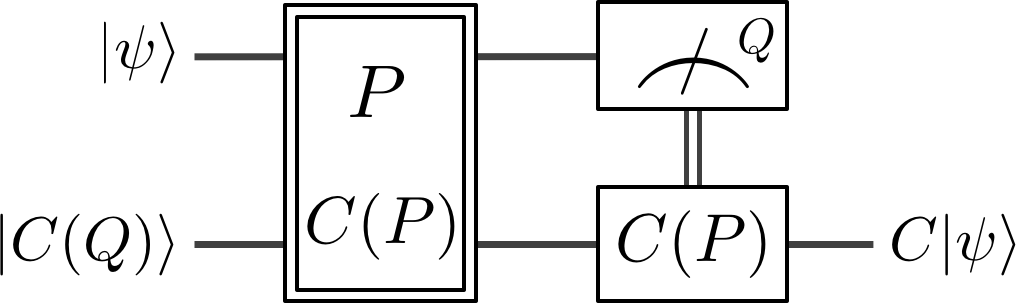}
\caption{\label{lattsurg_1bittele} A family of circuits for teleporting Clifford gates. One may choose any two (different) single-qubit Paulis $P,Q$ and any single-qubit Clifford $C$ (including identity). Then, let $C(R)=CRC^\dag$ and also let $\ket{R}$ be the $+1$ eigenstate of single-qubit Pauli $R$. In the diagram, the \edit{doubled} box represents a projector onto the $+1$ eigenspace of $P\otimes C(P)$. \edit{Such a projector is implemented by measuring $P\otimes C(P)$ \cite{Horsman2012} and then applying a Pauli correction in the case of measuring $-1$. In this case the correction would be $C(Q)$ on the lower block.}}
\end{figure}

\begin{figure}
\hspace*{0.1in}
\includegraphics[width=0.9\columnwidth]{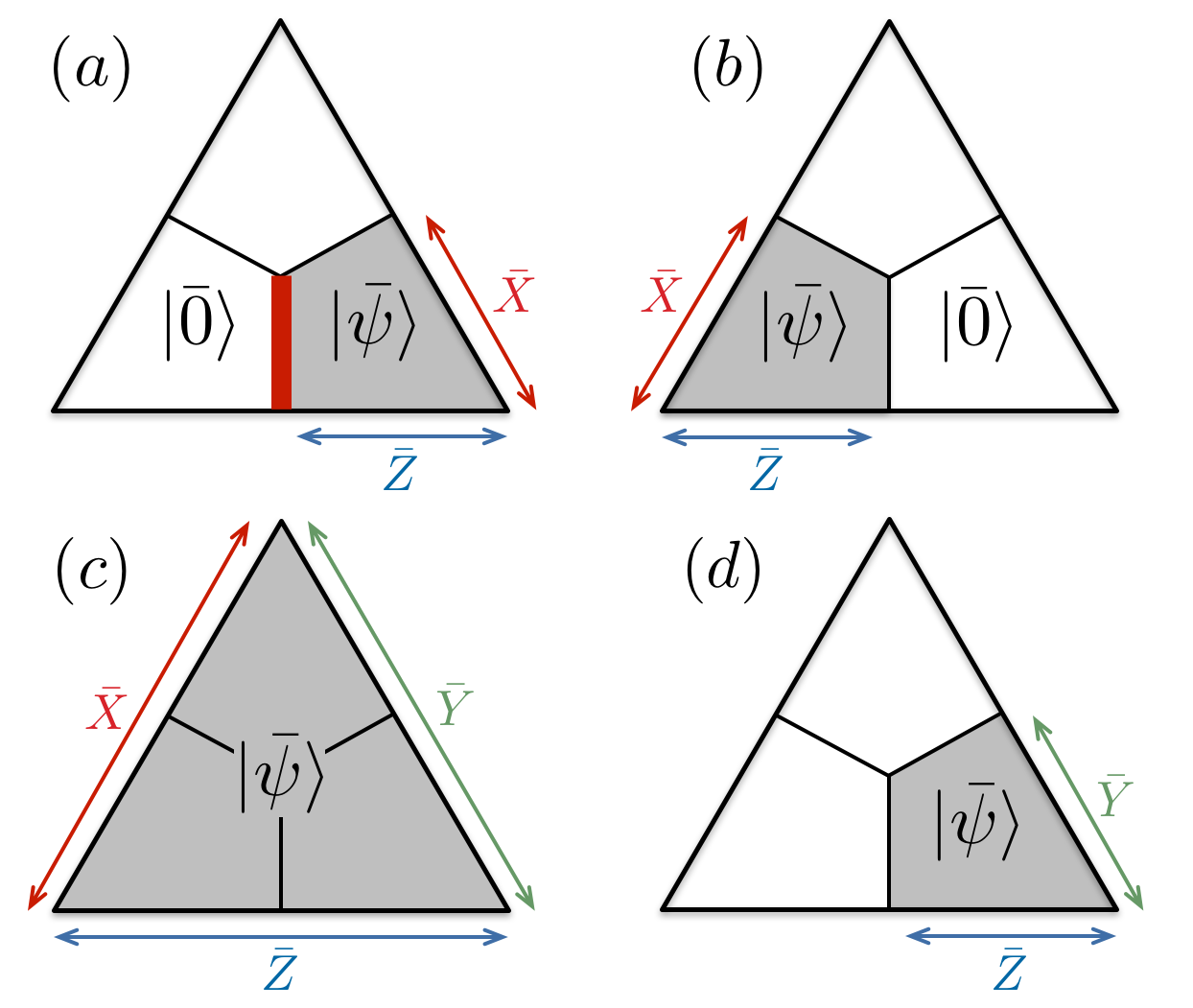}
\caption{\label{lattsurg_surfcodeSinv} Performing $S^\dag$ on a $d\times d$ surface code patch embedded in $1/3$ of a triangular patch as in Fig.~\ref{lattsurg_3layouts}. Colored arrows represent the directions of logical Paulis $\bar X$ (red), logical $\bar Z$ (blue), and logical $\bar Y$ (green). In (a), we prepare an ancilla in $|\bar0\rangle$ and project onto the $+1$-eigenspace of $\bar X_A\bar X_D$ of the ancilla and the data (thick, red line). In (b), we project what was the original data sector to $\ket{\bar0}$, completing the 1-bit teleportation (see Fig.~\ref{lattsurg_1bittele}) of $\ket{\bar\psi}$ to the ancillary sector. In (c), we extend the surface code to the full triangle code of distance $2d+1$, using Fig.~\ref{lattsurg_cornercreation}, then, in (d), shrink the code back down to the original sector. In the end, $\bar Y$ is located on the edge that originally held $\bar X$. Thus, $\bar Y\rightarrow\bar X$ and $\bar Z\rightarrow\bar Z$, so we have implemented $\bar S^\dag$.}
\end{figure}

The method of performing single-qubit gates is shown in Fig.~\ref{lattsurg_surfcodeSinv}. Through a composition of 1-bit teleportation (Fig.~\ref{lattsurg_1bittele}) and code conversion (Fig.~\ref{lattsurg_cornercreation}), we can place $\bar Y$ on the edge of the surface code. Enforcing the relabeling $\bar Y\rightarrow\bar X$, while keeping $\bar Z$ fixed performs the $\bar S^\dag$ gate. If we instead performed 1-bit teleportation into the top sector of Fig.~\ref{lattsurg_surfcodeSinv}(b), we perform the gate $\bar H\bar S^\dag\bar H$, completing a set of single-qubit Cliffords. We can also get $\bar H$ directly by the composition of three 1-bit teleportations to move $\ket{\bar\psi}$ around the twist (in either direction). \edit{These operations on patch-encoded qubits are analogous to how hole-encoded qubits are braided with a twist (here the triangle's center) to produce logical gates \cite{Brown2016}}.

The fact that we can perform the whole Clifford group on surface codes within a triangular patch layout is quite relevant to surface code research. Indeed, besides state distillation, there is only one other method we are aware of for performing $S$ on planar surface code patches \cite{Moussa2016}, and it effectively involves conversion to the color code through folding \cite{Kubica2015b}. It seems the triangle code offers a more natural approach, albeit with a not insignificant cost of $3d^2+O(d)$ physical qubits per logical.

Lower overhead computation can be achieved by using the triangle code as the resting code. Our second design Fig.~\ref{lattsurg_3layouts}(b) uses the \edit{BC} scheme of initialization and measurement. In this case, whenever we want to measure or initialize a patch we require ancilla qubits extending outside of the triangular region of the patch (see Sec.~\ref{sec:init_meas}). Thus, we require a neighboring ancilla patch empty of data. Moreover, the ancilla must be adjacent to the side containing the logical Pauli that we want to measure or initialize an eigenstate of. This influences all of our gate designs that use specially prepared ancilla patches.

\begin{figure}
\hspace*{0.15in}
\includegraphics[width=0.85\columnwidth]{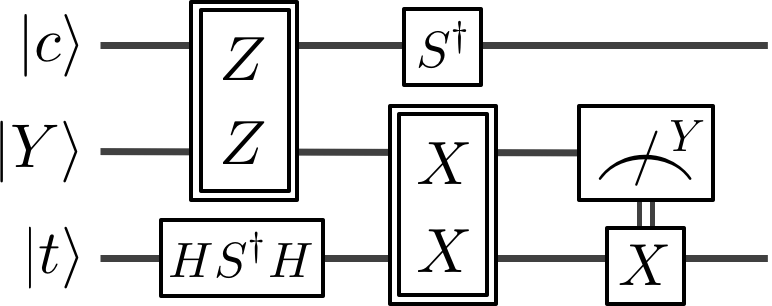}
\caption{\label{lattsurg_cnot} A circuit performing CNOT from $\ket{c}$ to $\ket{t}$ in the spirit of \cite{Horsman2012}. The $ZZ$ and $XX$ \edit{doubled} boxes are projectors onto the $+1$ eigenspace of those operators. We use this design specifically for computation in the layout of Fig.~\ref{lattsurg_3layouts}(b), where preparing and measuring the ancilla in the $Y$-basis is most natural, because that is the side of the ancilla patch that is not adjacent to data.}
\end{figure}

\begin{figure}
\hspace*{0.05in}
\includegraphics[width=0.95\columnwidth]{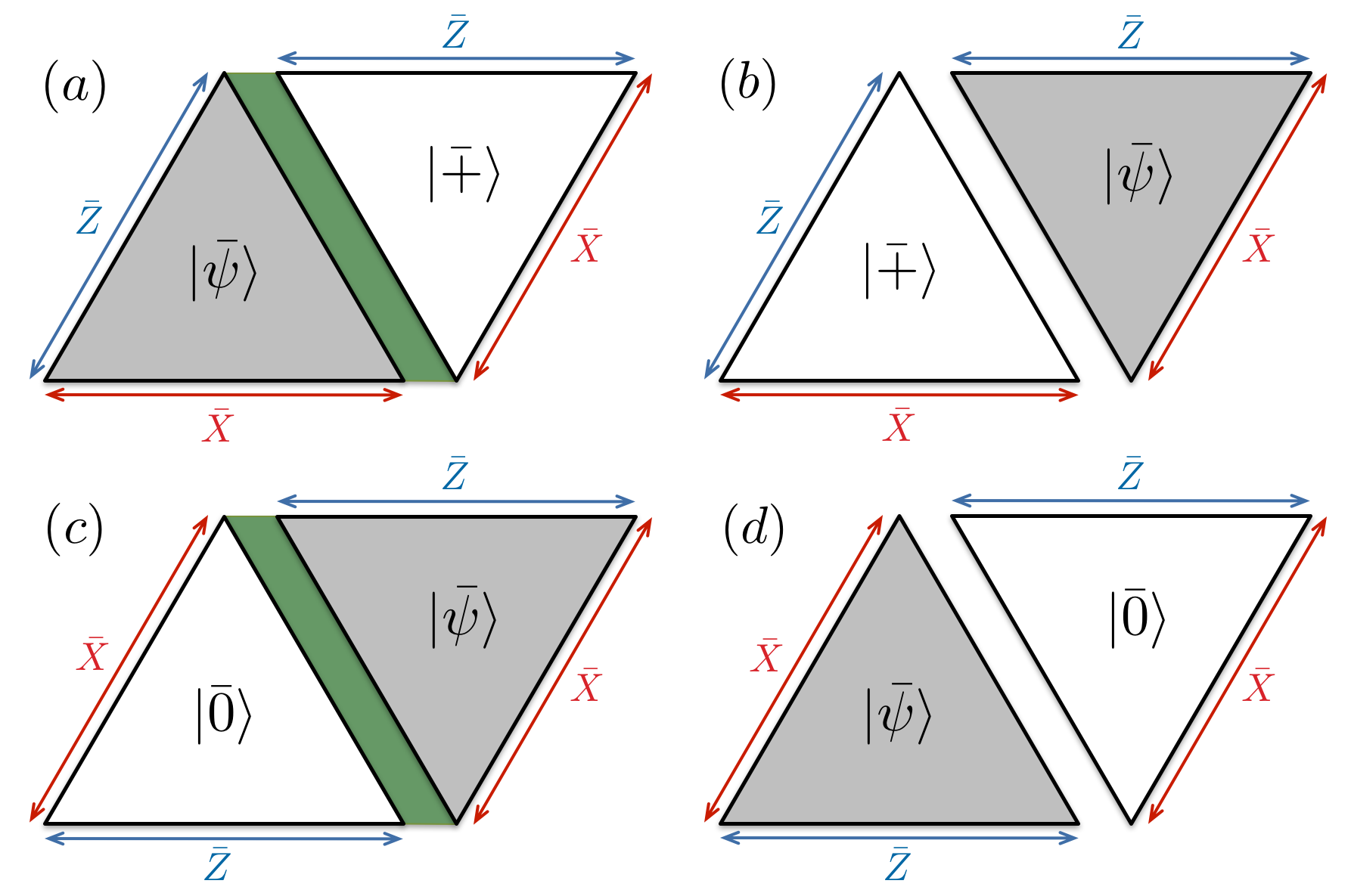}
\caption{\label{lattsurg_reorient} Reorienting a triangular patch by a reflection using a composition of two 1-bit teleportations Fig.~\ref{lattsurg_1bittele}. Colored arrows indicate the $\bar X$ (red) and $\bar Z$ (blue) sides of a patch, and green parallelograms \edit{between triangles in (a) and (c)} indicate projection onto $\bar Y\bar Y$. Notice that going from (b) to (c) is simply a relabeling of the ancilla --- $\ket{\bar+}$ oriented as in (b) is the same state as $\ket{\bar0}$ oriented as in (c). It is possible to view this process as a non-reorienting Hadamard (implemented by 1-bit teleportation with $C=H$ in Fig.~\ref{lattsurg_1bittele}) followed by a relabeling Hadamard (choosing to treat $\bar X$ as $\bar Z$ and vice versa), a composition which implements logical identity but changes orientation.}
\end{figure}

For instance, a projector circuit for CNOT is shown in Fig.~\ref{lattsurg_cnot}. To use this diagram with \edit{BC} initialization and measurement necessitates two ancilla patches -- (1) an ancilla patch adjacent to both control and target and (2) a second ancilla patch adjacent to the first -- and, moreover, that the data patches serving as control and target have their $\bar Z$ and $\bar X$ sides adjacent to the first ancilla, respectively. The layout of Fig.~\ref{lattsurg_3layouts}(b) guarantees the availability of the two ancillas. To ensure that the control and target patches are correctly oriented, notice that reorienting data patches can be done using 1-bit teleportation in the procedure prescribed by Fig.~\ref{lattsurg_reorient}. When Fig.~\ref{lattsurg_3layouts}(b) is tessellated, there is just enough room to reorient any data patch however we like, though not necessarily in parallel with some of the nearest other data patches. Both the reorientation and the CNOT itself take $O(d)$ time.

Single-qubit Clifford gates on the Fig.~\ref{lattsurg_3layouts}(b) layout can be done by simply relabeling sides of the code, treating, for instance, what was the $\bar Z$ side as $\bar X$ and vice versa. This method of single-qubit gates directly exploits the symmetry of the triangle code --- whatever subsequent lattice surgery might be done with one side can equally well be done with another. Notice that for this to work in concert with two-qubit gates, it is crucial that we have the aforementioned reorienting protocol.

Our final layout, \edit{using CS initialization and measurement}, is shown in Fig.~\ref{lattsurg_3layouts}(c). This approach uses $d$ ancilla qubits positioned between triangles \edit{to build CAT states}. In this scenario, logical initialization and measurement can be done without intruding upon neighboring patches, making gates quite simple conceptually. Fig.~\ref{lattsurg_cnot} suffices for coupling neighboring patches, Fig.~\ref{lattsurg_reorient} for reorienting them, and relabeling sides for single-qubit Clifford gates. The ancilla qubits do not get in the way of lattice surgery since we can always extend a triangular patch through them. The downside is the $O(d^2)$ time it takes to initialize and measure ancillas for a CNOT gate (and for reorientation).

\section{The smallest triangle code, code comparison, and logical tomography}\label{sec:distance_3}
Our goal in this section is to develop circuits for syndrome extraction, initialization, measurement, and gates on the smallest, fully fault-tolerant member of the triangle code family.\footnote{The 7-qubit, distance-3 triangle code was actually the first member of the family discovered by Andrew Cross at IBM T.J.~Watson through enumeration of 7-qubit stabilizer codes \cite{Cross2016}.} We view this as a worthy pursuit because of the qubit savings over the smallest, fully fault-tolerant surface code -- 13 versus 17. We will also see that the triangle code provides other advantages, such as the ability to perform tomography on the encoded logical qubit, also with only 13 qubits. We calculate pseudothresholds for our circuits, and find them comparable to the surface code, and better than other small fault-tolerant designs such as the color code. Thus, even though we do not claim our circuits are optimal in terms of depth or pseudothreshold, they at least show by construction that the smallest triangle code has the potential for a small demonstration of fault-tolerance.

Our first task is to create a circuit for syndrome extraction on the distance-3 triangle code. As mentioned in Sec.~\ref{sec:construction}, we can check by enumeration that this is impossible if we make the harshest demands on space and time requirements --- just 6 ancillas and 6 timesteps are not sufficient for error-correction on the distance-3 triangle code. Alternatively, the design of Fig.~\ref{circ_synextract} implies 8 ancillas (a total of 15 qubits) and 7 timesteps are sufficient to perform syndrome extraction for distance three. We now discuss a circuit identity that we can use to reduce those 8 ancillas to 6 (a total of 13 qubits), but with 8 timesteps.

\begin{figure}
\includegraphics[width=\columnwidth]{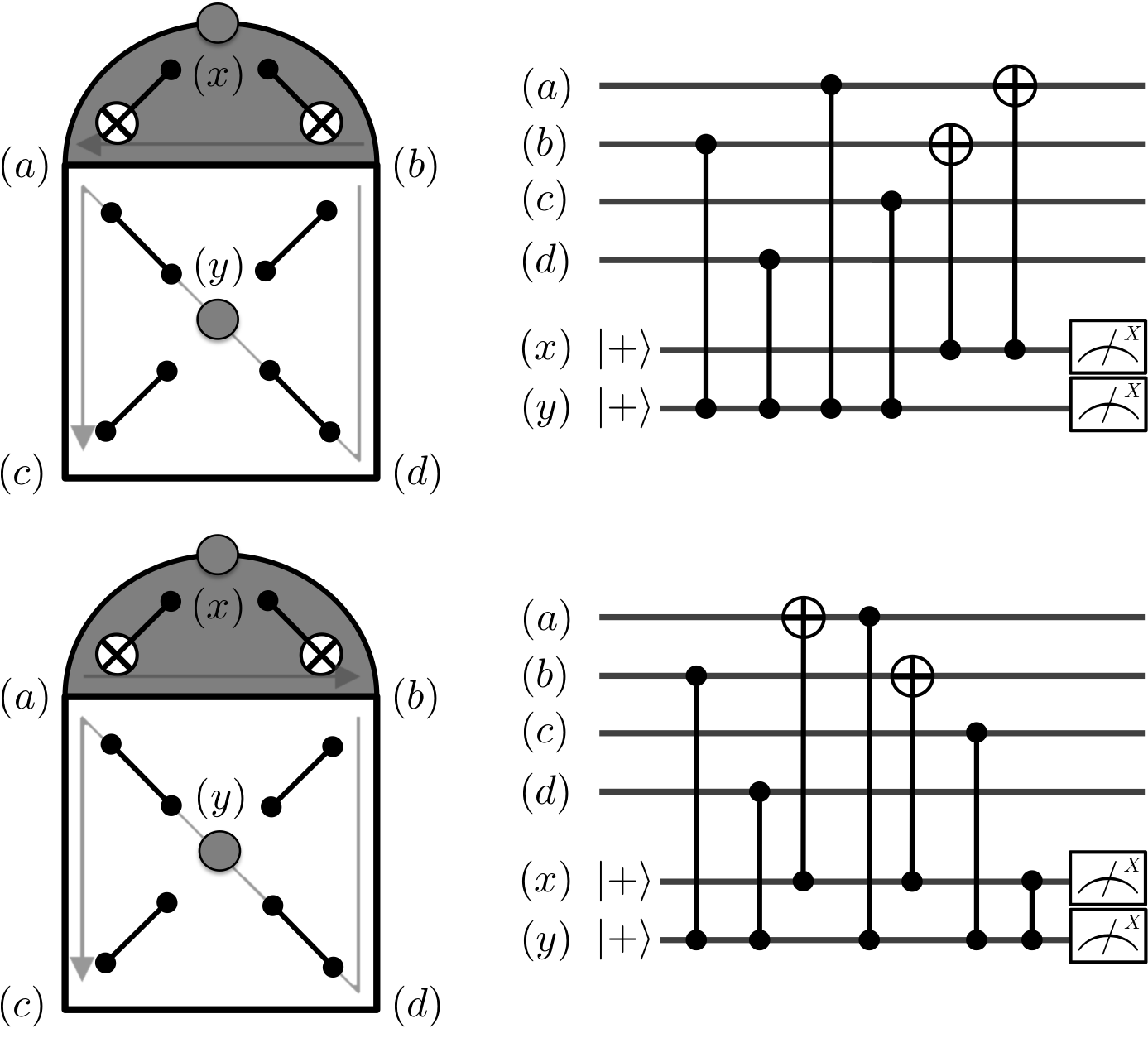}
\caption{\label{circ_outoforder} A plaquette stabilizer and loop undergoing syndrome extraction. The top image and circuit shows the conventional ordering, where coupling to the loop occurs completely after coupling to the plaquette. The lower circuit is equivalent (by circuit identities), but here coupling to the loop is interwoven with coupling to the plaquette. This rewriting, however, makes a significant difference to fault-tolerance, because the lower circuit can detect hook errors (e.g.~an $X$ occurring partway on qubit $(x)$ will propagate a $Z$ onto qubit $(y)$ flipping its measurement) while the upper cannot. }
\end{figure}

The idea is illustrated in Fig.~\ref{circ_outoforder}, where extraction of a loop's syndrome is \emph{interwoven} with extraction of the neighboring plaquette's syndrome. This ordering necessitates a CZ gate between the loop and plaquette ancillas before measuring, but that same CZ gate also endows the loop ancilla with the ability to detect hook errors caused by failure of the plaquette ancilla. Ideally, this CZ gate can be done directly between the ancillas, though this demands high connectivity (degree-5) for the plaquette ancilla. If degree-4 connectivity is desired, a depth-2 ``cascade" of CNOT and CZ gates can be used to perform the CZ gate between ancillas using a data qubit as an intermediary. This is still fault-tolerant (data qubits are not coupled to one another), but will lower the threshold due to larger circuit depth.

\begin{figure}
\hspace*{0.25in}
\includegraphics[width=0.75\columnwidth]{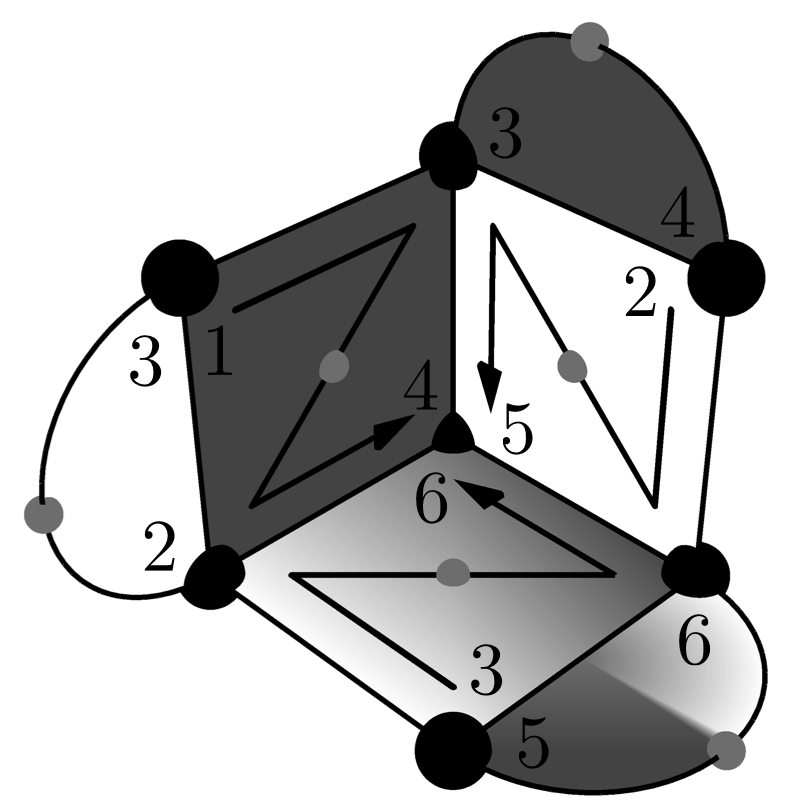}
\caption{\label{circ_trianglecode3} Proposed syndrome extraction for the distance-3 triangle code using the minimal number of ancillas. Numbers indicate the timesteps in which coupling the ancilla to data occurs. Extraction from the $X$-type and $Z$-type plaquettes is interwoven with extraction on their neighboring loops. If the CZ gate between interwoven ancillas can be done in one timestep, then this whole circuit takes 8 timesteps (including preparation and measurement of ancillas) to extract the complete syndrome once. However, two subsequent extractions can be performed in 15 timesteps.}
\end{figure}

Using interwoven extraction on two plaquettes instead of 2-CATs provides a circuit for distance-3 syndrome extraction with only 6 ancillas, Fig.~\ref{circ_trianglecode3}. For larger distance triangle codes, we leave the question open as to whether a similarly interwoven circuit suffices for full-distance extraction. In any case, in the large distance limit, a two-qubit savings over Fig.~\ref{circ_synextract} is less impactful.

In Tab.~\ref{d3_smallFTcompare}, we show a comparison of small distance-3 fault-tolerant designs by overhead and pseudothreshold for syndrome extraction. These designs were chosen as they satisfy reasonable criteria for a near-term fault-tolerant qubit: all use less than 20 qubits including ancillas, the connectivity between qubits required is low, and no design uses postselection to create ancillas during syndrome extraction. The triangle code performs remarkably well in the comparison, saving two or four qubits over the surface code with a drop in pseudothreshold of only $3\%$ or $14\%$, respectively. In Appendix~\ref{app_pseudothresh}, we provide more detail on the circuits and the exact-counting procedure used to determine these pseudothresholds.

\setlength\extrarowheight{4pt}
\begin{table*}
\setlength{\tabcolsep}{0.35em}
\begin{tabular}{|c||c|c|cc|c|cc|cc|}
\hline
 \text{Code} & \text{Data} & \text{Anc.} & \text{Qubits} & \text{($\#$/surf.)} & \text{Max deg.} & \text{Lower pseudoth.} & \text{($\#$/surf.)} & \text{Upper pseudoth.} & \text{($\#$/surf.)} \\
\hline\hline
 \text{5-qubit} & 5 & 3 & 8 & \text{(.47)} & 3 & $1.02\times10^{-5}$ & \text{(.05)} & $1.05\times10^{-5}$ & \text{(.04)} \\
\hline
 \text{5-qubit} & 5 & 6 & 11 & \text{(.65)} & 3 & $2.55\times10^{-5}$ & \text{(.14)} & $2.69\times10^{-5}$ & \text{(.10)} \\
\hline
 \text{Color} & 7 & 6 & 13 & \text{(.76)} & 3 & $3.47\times10^{-5}$ & \text{(.19)} & $3.77\times10^{-5}$ & \text{(.15)} \\
\hline
 \text{5-qubit$^*$} & 5 & 12 & 17 & \text{(1.0)} & 4 & $5.20\times10^{-5}$ & \text{(.28)} & $5.70\times10^{-5}$ & \text{(.23)} \\
\hline
 \text{Color$^*$} & 7 & 12 & 19 & \text{(1.1)} & 6 & $5.80\times10^{-5}$ & \text{(.31)} & $6.65\times10^{-5}$ & \text{(.26)} \\
\hline
 \text{Surface} & 9 & 4 & 13 & \text{(.76)} & 4 & $7.85\times10^{-5}$ & \text{(.43)} & $9.40\times10^{-5}$ & \text{(.38)} 
\\
\hline
 \text{Bacon-Shor} & 9 & 8 & 17 & \text{(1.0)} & 3 & $8.98\times10^{-5}$ & \text{(.49)} & $1.07\times10^{-4}$ & \text{(.43)} \\
 \hline
 \text{Triangle} & 7 & 6 & 13 & \text{(.76)} & 4 & $1.05\times10^{-4}$ & \text{(.57)} & $1.22\times10^{-4}$ & \text{(.49)} \\
\hline
 \text{Triangle} & 7 & 6 & 13 & \text{(.76)} & 5 & $1.57\times10^{-4}$ & \text{(.86)} & $1.92\times10^{-4}$ & \text{(.77)} \\
\hline
 \text{Triangle} & 7 & 8 & 15 & \text{(.88)} & 4 & $1.76\times10^{-4}$ & \text{(.97)} & $2.23\times10^{-4}$ & \text{(.90)} \\
\hline
 \text{Surface} & 9 & 8 & 17 & \text{(1.0)} & 4 & $1.82\times10^{-4}$ & \text{(1.0)} & $2.47\times10^{-4}$ & \text{(1.0)} \\
\hline
\end{tabular}
\caption{\label{d3_smallFTcompare} A comparison of small, fully fault-tolerant designs across qubit count, maximum degree connectivity required, and pseudothreshold (under depolarizing noise) for the exREC of a transversal single-qubit gate (e.g.~identity). In parentheses, quantities are written as a fraction of the corresponding surface code quantity. The pseudothreshold upper bounds apply only to the particular circuits and error model considered, but are useful insofar as they can definitively prove separations between the logical error rates of different designs. Starred (*) designs use connectivity that is not planar. All designs are capable of logical qubit tomography with just the qubit resources listed, except for the Bacon-Shor and surface code designs, which are lacking initialization and measurement in the $Y$-basis. Although the surface code does possess the largest provable pseudothreshold in our study, it does not do so by much, and a fewer-qubit triangle code design may prove to be a more practical small logical qubit.}
\end{table*}

It is also important that we be able to initialize and measure logical states in the distance-3 triangle code. It turns out that each of the triangle code designs from Tab.~\ref{d3_smallFTcompare} can perform initialization and measurement using only the qubits provided. The idea is quite simple: to measure $\bar Z$ we need only use the mixed-type plaquette ancilla, because $\bar Z=Z_{(1,0,0)}Z_{(0,0,0)}Z_{(0,1,0)}$ is supported on qubits adjacent to it. Performing a measurement of $\bar Z$ before a round of syndrome extraction, and repeating the process at most three times, will successfully measure $\bar Z$. Since $\bar X$ and $\bar Y$ are also localized around single plaquettes, similar procedures work for measuring them.

We show pseudothrsholds for measurement of distance-3 triangle code designs in Tab.~\ref{d3_meascompare}. They are lower than the pseudothresholds of transversal gates shown in Tab.~\ref{d3_smallFTcompare}. This is to be expected from the slightly larger circuit depth, caused by having to measure $\bar Z$. Although a lower measurement threshold is not disastrous (measurement occurs at most once in a single-qubit experiment), we also show that, if necessary, the pseudothreshold can be raised by adding one additional qubit. Details are provided at the end of Appendix~\ref{app_pseudothresh}.

\begin{table}
\hspace*{0.05in}
\begin{tabular}{|c|c|c|c|}
\hline
Qubits & Max.~Deg. & Lower & Upper\\
\hline\hline
13 & 4 & $9.84\times10^{-5}$ & $1.18\times10^{-4}$ \\
\hline
15 & 4 & $1.07\times10^{-4}$ & $1.27\times10^{-4}$ \\
\hline
13 & 5 & $1.15\times10^{-4}$ & $1.39\times10^{-4}$ \\
\hline
14 & 5 & $1.84\times10^{-4}$ & $2.65\times10^{-4}$ \\
\hline
\end{tabular}
\caption{\label{d3_meascompare} A comparison of pseudothresholds for measurement exRECs of $d=3$ triangle code designs. The first three designs match the three from Tab.~\ref{d3_smallFTcompare}, and have lower pseudothresholds for measurement than they do for gates. The last design is the 13-qubit, degree-5 design augmented by an extra qubit for measuring the logical operator (see Appendix~\ref{app_pseudothresh}, Fig.~\ref{app_measplusone}).}
\end{table}

\begin{figure}
\hspace*{0.25in}
\includegraphics[width=0.8\columnwidth]{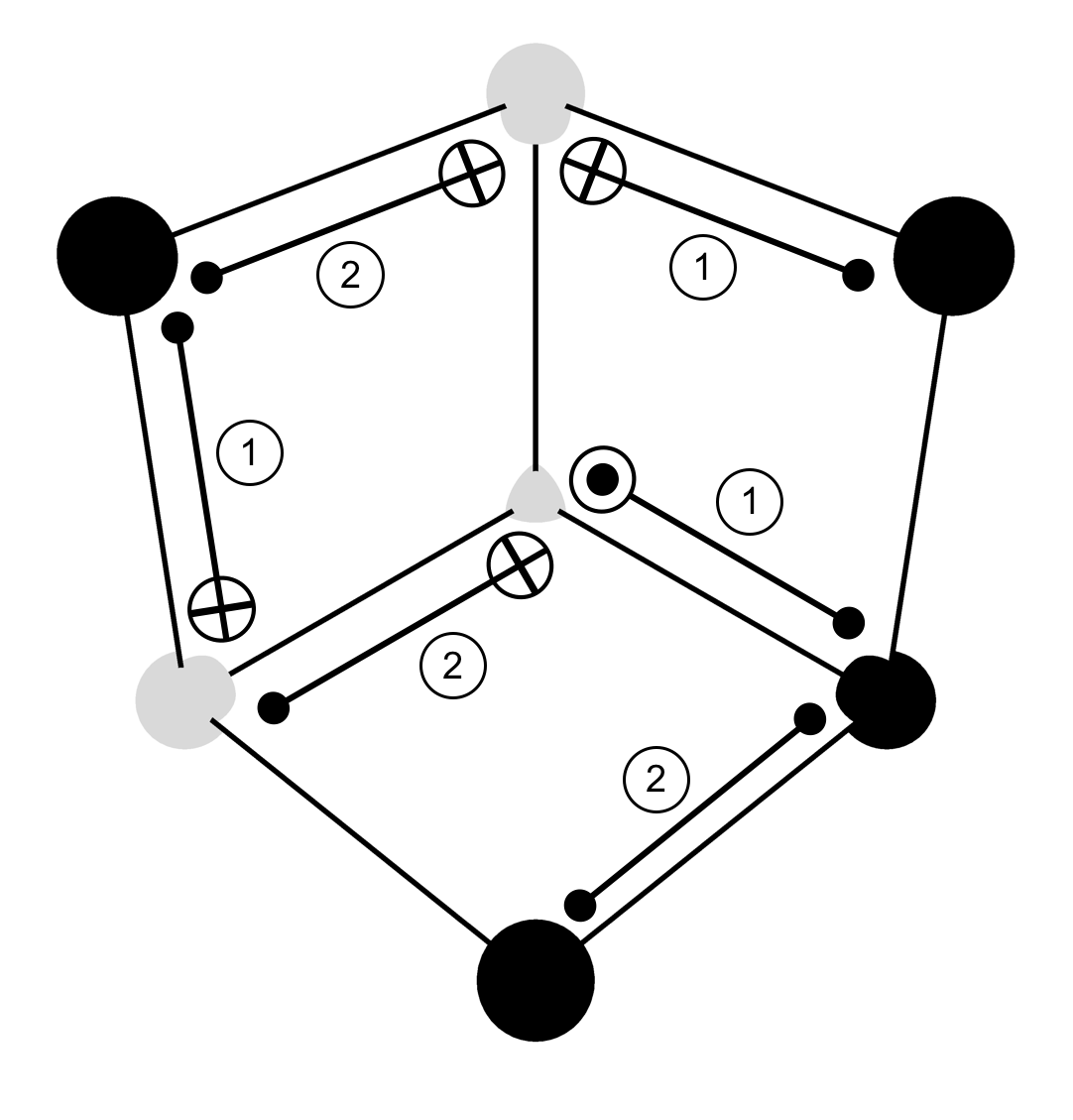}
\caption{\label{circ_triinitialization2} Creating the $|\bar0\rangle$ state in the distance-3 triangle code can be done with this depth-2 circuit. Light qubits should be prepared in $|0\rangle$ and dark qubits in $|+\rangle$. Numbered buttons indicate the timesteps in which gates should be performed. The notation $\bullet\text{---}\odot$ indicates a controlled-$Y$ gate. Although the gates are pictured acting between data qubits, if the connectivity of the implementation does not cooperate, they can also be indirectly implemented by CNOT cascades through the ancilla qubits (not shown).}
\end{figure}

The most reliable way of initializing the triangle code is through a non-fault-tolerant circuit followed by postselection. In Fig.~\ref{circ_triinitialization2}, we show a depth-2 circuit that can prepare $|\bar0\rangle$ non-fault-tolerantly. Similar circuits exist for preparing eigenstates of $\bar X$ and $\bar Y$. To guarantee fault-tolerance, Fig.~\ref{circ_triinitialization2} should be followed by postselection on trivial measurements of the syndrome and of $\bar Z$. Since this postselection is done only at the beginning of an experiment, we view it as not too restrictive for a small fault-tolerant design.

The logical operations we have described here, preparation and measurement in any Pauli basis and error-correction, serve to implement essentially any interesting single-logical-qubit experiment on the distance-3 triangle code. For instance, process tomography of logical identity $\bar I$ can be performed by preparing logical Pauli eigenstates and then performing logical measurement in a Pauli basis. Tomography of $\bar I^k$ is the same with more rounds of error-correction between preparation and measurement. Single-qubit logical Cliffords are essentially rotations or reflections of the triangle, which can be done without need for physical Cliffords. If desired for tomographic purposes, physical Cliffords can be applied to transform to a locally equivalent triangle code without significantly affecting the circuits for syndrome extraction or measurement. A special case of this is a transversal, order-3 Clifford gate $\overline{SH}$: perform $H$ on all qubits at positions $x>0$ or $x=0,z>0$ and $SH$ on the central qubit. The resulting code is the same as the initial triangle code rotated $120^\circ$ counterclockwise. Since the logical Pauli operators lie along the triangle's sides, they are cyclically permuted by this operation. 

\section{Conclusion}\label{sec:conclusion}
It is important to keep in mind that, with fault-tolerant quantum computing experiments still in the nascent stage, it is difficult to know what constitutes an ideal scheme for local, planar computation. Instead, it is crucial to develop a broad array of tools for fault-tolerance, and hope to meet experiments somewhere in the middle sometime in the future. The triangle code, with its accompanying syndrome extraction circuits and lattice surgery methods, is another tool in this toolkit.

Indeed, in addition to developing planar computation with patches of triangle code, we have also used the notion of a triangle code and a twist to improve surface codes, creating a simple surgery to implement $S$ without state distillation. This is some indication that thinking in terms of triangle codes is worthwhile, even if the triangle code itself is not the resting code of some topologically fault-tolerant architecture. To expedite logical gates, future quantum processors might rather be a tessellation of triangles than a grid.

It would be interesting to generalize the triangle codes to patches with more than one central twist, while still using only weight-4 plaquette stabilizers. One possibility is to take the 3-dimensional visualization of Fig.~\ref{intro_triangleplanar}(a) and attach more planes of surface code, creating mixed-type stabilizers and twists where necessary. As one example of this process, we find a topologically closed 2-dimensional surface (conveniently visualized as the surface of a cube with $d$ qubits per edge) containing 8 twists and encoding two logical qubits with distance $d$. This code is topologically distinct from the toric code. Determining properties or the usefulness of generalized twist-containing patches such as this, and maybe different topologies encoding even more qubits, is a possible direction of future research.

\section*{Acknowledgement}
We especially thank Andrew Cross for encouraging us to study this family of codes, for many helpful discussions, and for performing the topological threshold calculation. This work would not have been possible without his guidance. In addition, we thank Sergei Bravyi, Isaac Chuang, and Jay Gambetta for discussions and comments on the manuscript. T.J.Y. acknowledges the hospitality of IBM T.J.~Watson where much of this work was done. T.J.Y. also thanks the National Defense Science and Engineering Graduate (NDSEG) fellowship and NSF RQCC Project No.~1111337 for support.

\appendix

\section{Memory threshold}\label{app_thresh}
In this appendix, we first prove that the family of triangle codes possesses a fault-tolerance threshold in the asymptotic limit, and then discuss our simulation for estimating it. The error model is not based on a circuit for syndrome extraction. Single-qubit errors on the data occur with probability $p$, and a syndrome bit is flipped with probability $q$. As is common in these matters, the estimated threshold is much better than the proven one.

Our argument closely follows that of Dennis et.~al.~\cite{Dennis2002} who prove a threshold for the toric code. The important differences are (1) our code is not on a closed topological surface and (2) there is a single decoding graph, rather than two separate decoding graphs for $X$ and $Z$ errors (because the triangle code is not CSS). As a consequence of the latter property, while we can prove a threshold with magnitude similar to that for the toric code in the case of uncorrelated $X$ and $Z$ errors, we can prove only a lower threshold (by about a factor of 36) in the case of correlated errors (i.e.~single-qubit depolarizing noise).

\begin{figure}
\hspace*{0.12in}
\includegraphics[width=0.9\columnwidth]{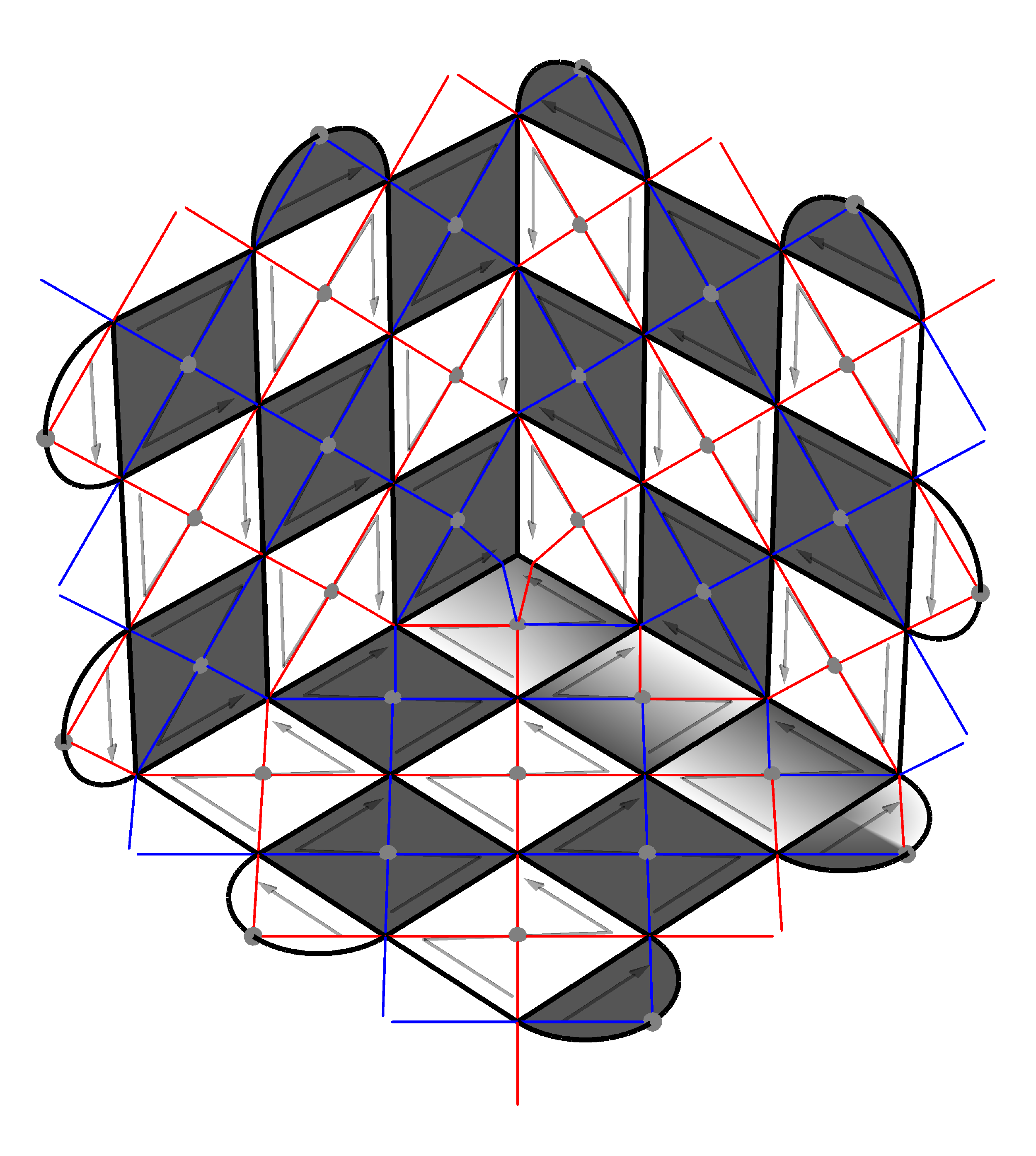}
\caption{\label{app_decodegraph} The decoding graph for a (single layer of) triangle code overlaid on the code. Nodes in the graph are located at each stabilizer (gray dots) and edges are colored blue or red. Some edges are connected to only one node (or alternatively are connected to a node at infinity). These are the boundary edges defined in the text. Also shown in the figure is the order of syndrome extraction as per Fig.~\ref{circ_synextract}.}
\end{figure}

The decoding graph is overlaid on the triangle code in Fig.~\ref{app_decodegraph}. To account for errors in measuring the syndrome, we envision a three-dimensional stack of these decoding graphs, with vertical edges connecting corresponding qubits between the layers. We call edges connected to only one node ``boundary edges" and notice that they can be grouped into three sets corresponding to the side of the triangle code stack they occupy.

Every set of errors $E$, in general a combination of data errors and syndrome errors, can be associated with a subset of edges $P$ of the decoding graph. The association is also true the other way: a subset of edges uniquely specifies a set of errors. Each data qubit has two associated edges, a blue edge and a red edge. The blue edge is in $P$ if and only if a $Z$ error occurred on that qubit, and the red edge is in $P$ if and only if an $X$ error did. A $Y$ error corresponds to both edges being in $P$. Vertical edges between layers correspond to bit flip errors on the syndrome. Using this correspondence we subsequently speak of errors as edges without distinction.

We can also relate logical errors to paths on the decoding graph. A logical error corresponds to a path of length at least $d$ (the code distance) going from one boundary edge to another boundary edge occupying a \emph{different} side of the triangle code. The path may traverse several layers before terminating on another side, though paths that do are strictly longer than length $d$.

After $T$ cycles of syndrome extraction, we have accumulated errors $E$ and would like to apply a recovery operation. A simple strategy, and that considered in \cite{Dennis2002}, is to apply a recovery $E_m$ with $H_m$ horizontal edges and $V_m$ vertical edges such that (1) $\partial E=\partial E_m$ and (2) $H_m\log\left(\frac{1-p}{p}\right)+V_m\log\left(\frac{1-q}{q}\right)$ is minimal. Here $\partial E$ for a set of edges $E$ is the boundary of $E$, the set of nodes connected to only one edge in $E$. Notice that (2) is defined implicitly for correcting uncorrelated $X$ and $Z$ errors, because it cares about the number of edges in $E_m$, rather than the number of affected qubits. We will change the recovery later to deal with correlations.

What is the probability that a path $P$ consisting of $H$ horizontal edges and $V$ vertical ones is contained in $E\cup E_m$? If $H_e$ and $V_e$ are the numbers of horizontal and vertical edges in $E$, then $P\subseteq E\cup E_m$ implies that $H_e+H_m\ge H$ and $V_e+V_m\ge V$. Accordingly,
\begin{align}
\left(\frac{p}{1-p}\right)^{H}\left(\frac{q}{1-q}\right)^{V}\ge&\left(\frac{p}{1-p}\right)^{H_m}\left(\frac{q}{1-q}\right)^{V_m}\\\nonumber&\times\left(\frac{p}{1-p}\right)^{H_e}\left(\frac{q}{1-q}\right)^{V_e}.
\end{align}
Using properties (1) and (2) of the recovery, we know that the product of the former two factors on the left side of the inequality is greater than the product of the latter two. This implies
\begin{equation}
\left(\frac{p}{1-p}\right)^{H_e}\hspace{-3pt}\left(\frac{q}{1-q}\right)^{V_e}\hspace{-4pt}\le\left(\frac{p}{1-p}\right)^{H/2}\hspace{-3pt}\left(\frac{q}{1-q}\right)^{V/2}
\end{equation}
The probability $E$ occurs is $\le\hspace{-4pt}p^{H_e}(1-p)^{H-H_e}q^{V_e}(1-q)^{V-V_e}$, and there are $\le\hspace{-4pt}2^{H+V}$ ways to distribute edges in $P$ between $E$ and $E_m$. So, we find the bound 
\begin{equation}
\text{Pr}\left[P\subseteq\hspace{-2pt}E\cup E_m\right]\le(4p(1-p))^{H/2}(4q(1-q))^{V/2}.
\end{equation}
Summing over the possible logical errors bounds the probability of a logical error.
\begin{equation}\label{error_bound}
\text{Pr}\left[\text{error}\right]\le\sum_{H,V}N_{H,V}(4p(1-p))^{H/2}(4q(1-q))^{V/2},
\end{equation}
where $N_{H,V}$ is the number of paths containing $H$ horizontal and $V$ vertical edges that represent logical errors. For instance, $N_{H,V}=0$ whenever $H<d$.

While $N_{H,V}$ is difficult to calculate for general $H\ge d$ and $V$, it is trivial to get a satisfactory bound. Along the boundary of the triangle code stack, which is $T$ layers tall, there are fewer than $T\times(3d)$ boundary edges on which to start a logical error path $P$. From there, at each node we can typically continue the path in one of at most five directions, three remaining in the same layer and two changing layers. The exception is at the central mixed-type node of any layer, where we have at most six choices. A very rough upper bound is then that there are $N_{H,V}\le T\times(3d)\times6^{H+V}$ logical error paths. Putting this together with Eq.~\eqref{error_bound}, we find that $\text{Pr}\left[\text{error}\right]$ vanishes in the limit $d\rightarrow\infty$ whenever $p,q< p_{\text{th}}$ for some $p_{\text{th}}$ that we have now shown is at least $0.7\%$ in the case of uncorrelated $X$ and $Z$ errors.

When $X$ and $Z$ errors are correlated, or, in other words, the data-qubit noise model is single-qubit depolarizing noise, then we should modify property (2) of the recovery to ($2'$) $Q_m\log\left(\frac{1-p}{p}\right)+V_m\log\left(\frac{1-q}{q}\right)$ is minimal. Here $Q_m$ is the number of \emph{qubits} affected by the recovery. This is not the same as the number of horizontal edges, because each qubit is associated with two edges, one blue and one red. However, using the same reasoning as above, we can argue that
\begin{equation}\label{error_boundQ}
\text{Pr}\left[\text{error}\right]\le\sum_{Q,V}N'_{Q,V}(4p(1-p))^{Q/2}(4q(1-q))^{V/2},
\end{equation}
where now $N'_{Q,V}$ is the number of paths on the decoding graph containing $Q$ qubits and $V$ vertical edges that also correspond to a logical error. Each qubit is associated with two edges, so $N'_{Q,V}\le N_{2Q,V}\le T\times(3d)\times6^{2Q+V}$. Accordingly, the threshold is $p'_{\text{th}}\ge0.019\%$ in the case of depolarizing noise.

In practice, the thresholds are higher than our simple analytic estimates. The corresponding numerical threshold estimates in Table~\ref{tab:phenom} were calculated in the following way. Storage noise is modeled by applying a channel to each data qubit prior to syndrome measurement. The noise channel is either a bit flip channel with error probability $p$ or the composition of bit flip and phase flip channels, each with error probability $p$. We do not consider depolarizing noise since error correction by reduction to minimum weight matching is not maximally fault-tolerant for the triangle code for that case (instead, minimum qubit matching as described just above Eq.~\eqref{error_boundQ} would be sufficient, but we make no claims about the efficiency of that). As in \cite{Dennis2002}, we consider two different syndrome measurement models. The ``ideal'' model applies memory noise, measures one error-free syndrome, and performs error correction. The ``noisy'' model simulates faulty syndrome measurement using a sequence of $d+1$ syndrome measurement rounds. In each of the first $d$ rounds, we apply memory noise, measure one error-free syndrome, and flip each syndrome bit with probability $p$. In the final round, we measure one error-free syndrome. The error correction is computed based on all of the syndrome outcomes.

An algorithm infers error corrections from syndrome outcomes by reduction to minimum weight perfect matching \cite{Dennis2002}. The edge weights for the matching algorithm are computed differently for the triangle and rotated codes. We use the Manhattan distance for the rotated codes and process bit flip errors and phase flip errors separately. For the triangle code, we construct the decoding graph shown in Fig.~\ref{app_decodegraph}, assign each edge the same weight, and compute the total weight of any path using Dijkstra's algorithm. For the ``noisy'' model, multiple copies of Fig.~\ref{app_decodegraph} are joined by edges corresponding to potential syndrome bit errors.

\begin{figure}
\includegraphics[width=3.5in]{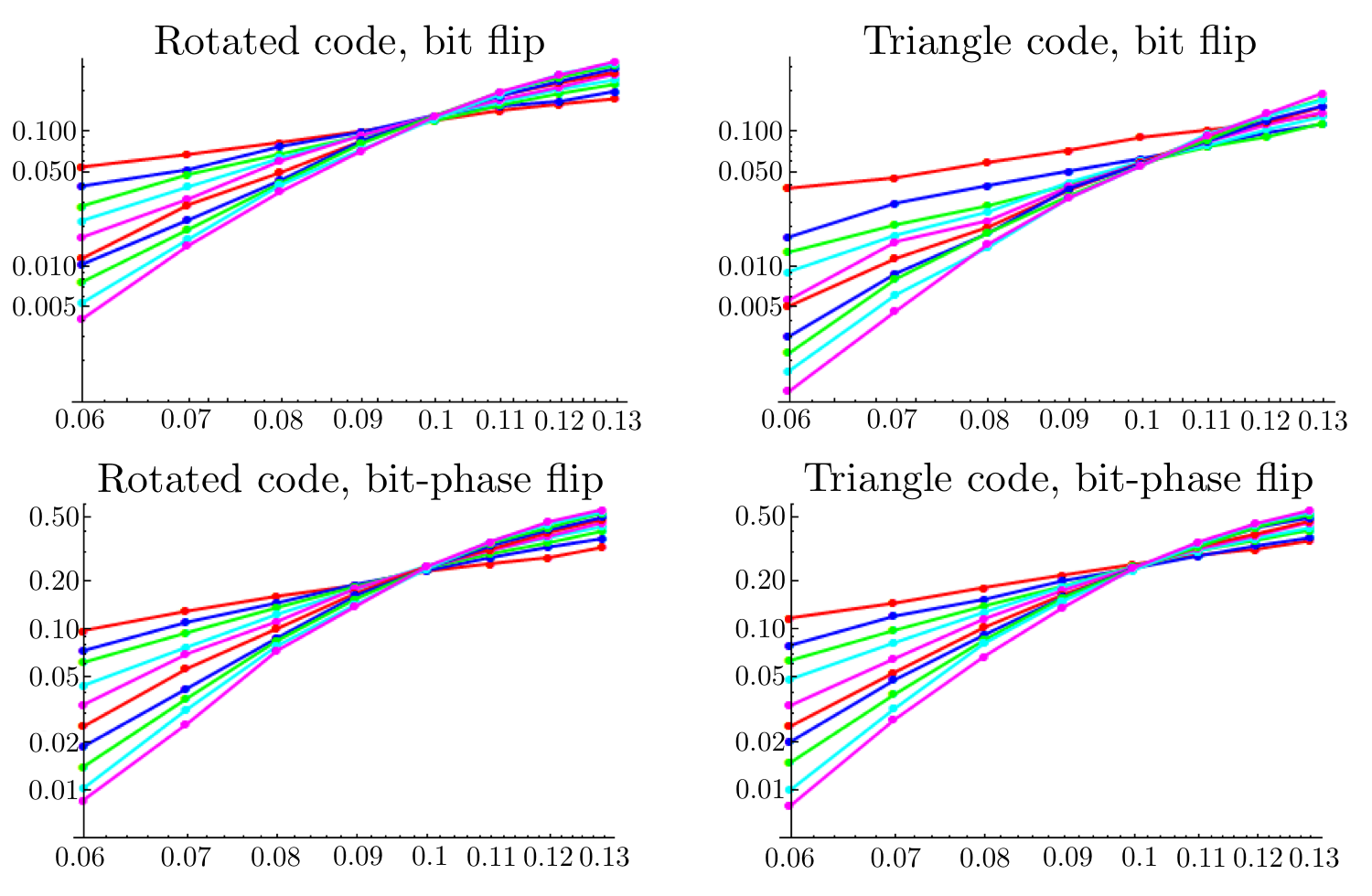}
\caption{Logical error rates for bit flip and bit-phase flip noise with ideal syndrome measurements suggest thresholds near $10\%$ for both the triangle and rotated codes. Odd code distances from $3$ to $21$ are shown, where each point corresponds to either 10,000 or 30,000 Monte-Carlo samples.\label{fig:idealSyndromePlot}}
\end{figure}

\begin{figure}
\includegraphics[width=3.5in]{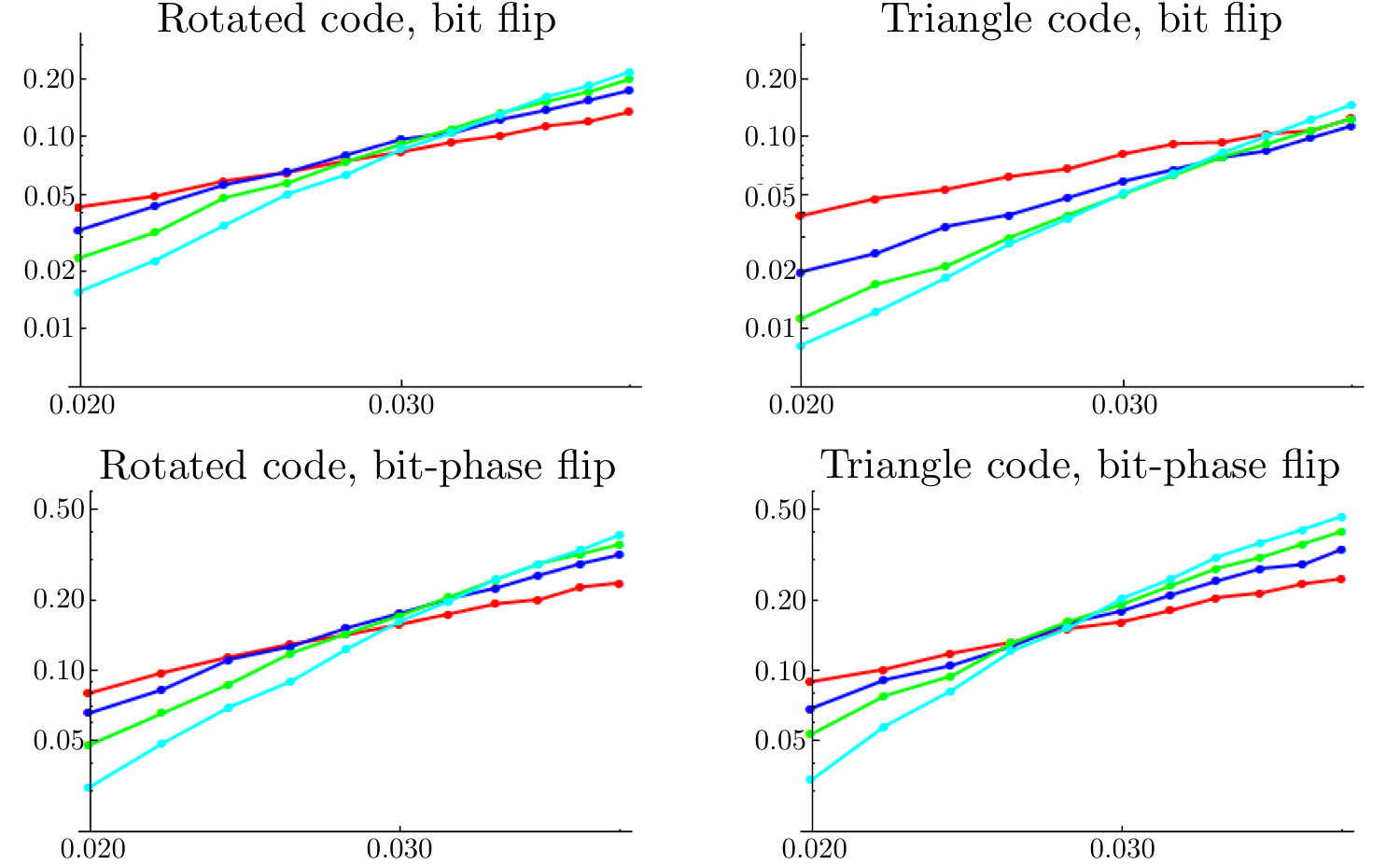}
\caption{Logical error rates for bit and bit-phase flip noise with noisy syndrome measurements suggest thresholds near $2.6\%$ for the triangle code and $3.2\%$ for the rotated code. Odd code distances $3$, $5$, $7$, and $9$ are shown, where each point corresponds to either 10,000 or 30,000 Monte-Carlo samples.\label{fig:noisySyndromePlot}}
\end{figure}

Fig.~\ref{fig:idealSyndromePlot} and Fig.~\ref{fig:noisySyndromePlot} show plots of all of the simulation results. It is worth noting that the error rate of the triangle code is generally much lower than that of the surface code in the presence of only bit flip noise, while it is close to the same in the case of bit-phase flip noise. This is roughly consistent, however, with our counts of lowest-weight logical operators in Tab.~\ref{logical_count}.

\begin{table}[t]
\setlength{\tabcolsep}{0.5em}
\begin{tabular}{|c|c|c|c|c|c|c|}
\hline
\multicolumn{2}{|c|}{}&\multicolumn{5}{|c|}{\# of weight $d$ logical strings}\\\hline
code&errors&d=3&5&7&9&11\\\hline
\multirow{3}{*}{surface}&X&8&52&296&1,556&7,768\\
&X,Z&16&104&592&3,112&15,536\\
&X,Y,Z&16&104&592&3,112&15,536\\\hline
\multirow{3}{*}{triangle}&X&3&20&95&546&2,583\\
&X,Z&14&86&476&2,462&12,164\\
&X,Y,Z&30&204&1,164&6,072&30,012\\
\hline
\end{tabular}
\caption{\label{logical_count} Counting the number of lowest-weight logical operators that can be created from physical errors of various types. While the triangle code has fewer lowest-weight logical operators that can be made from just bit or bit and phase errors, it has many more when correlated errors are allowed.}
\end{table}

\section{Effective distance of syndrome extraction}\label{app_synextract}
Our goal in this appendix is to (semi-rigorously) argue that the syndrome extraction circuit in Fig.~\ref{circ_synextract} has effective distance $d$. That is, there is no set of fewer than $d$ faults that can cause a logical error. Since Fig.~\ref{circ_synextract} is based on full-distance extraction for the surface code, we first review the argument in that case.

\begin{figure}
\includegraphics[width=\columnwidth]{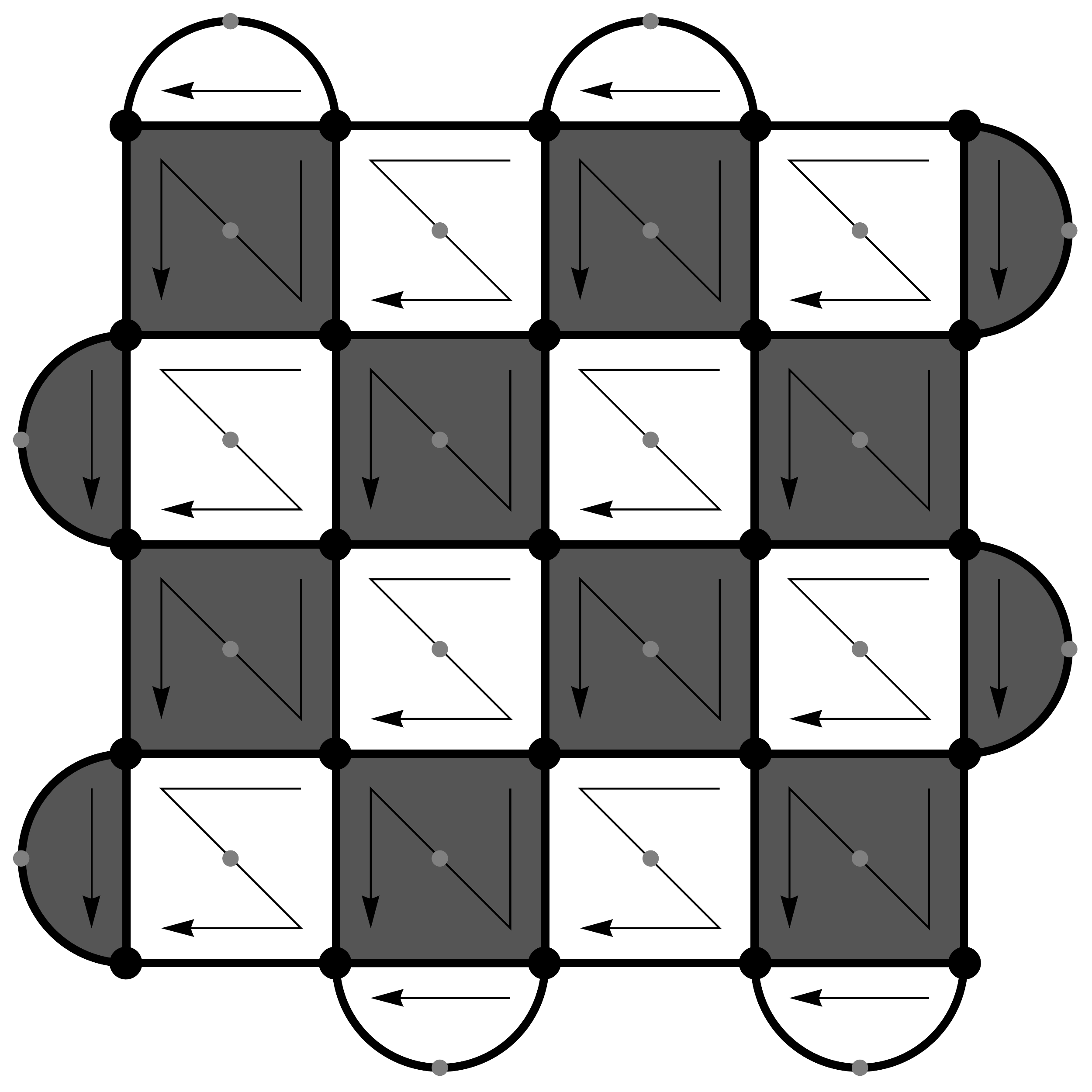}
\caption{\label{circ_surfacecode} The pattern for syndrome extraction in the surface code recommended in \cite{Tomita2014}. Notice that $Z$-type stabilizers should be measured in a pattern perpendicular to the $X$-type edges, so that a single hook error cannot spread $Z$ errors to two different rows of data qubits. Likewise, the $X$-type stabilizers are measured perpendicular to the $Z$-type edges.}
\end{figure}

Consider the syndrome extraction circuit indicated by Fig.~\ref{circ_surfacecode}. We aim to show that no set of $<d$ faulty circuit components can cause a logical failure. Because the surface code is CSS, we can consider $X$ and $Z$ errors separately, and refer to a $Y$ error as both an $X$ and a $Z$. A logical $\bar X$ is simply a string of $X$s between the $X$-type edges. This implies that, to form $\bar X$, we need at least one $X$ error per column of data qubits. However, we can now check that every faulty circuit component leads to $X$ errors in at most one column. In fact, only ``hook'' faults (a failure of the second or third gate coupling an ancilla to data) can even leave more than one error on the data. If the syndrome extraction for stabilizer $s$ couples to qubits $a,b,c,d$ in that order, then the errors resulting from hooks take the form $*_cs_d$ and $s_a*_b$, where $*$ is any Pauli and $s_j$ is the $j^{\text{th}}$ Pauli of $s$. Hence, the only hooks that can leave two $X$ errors on the data are from measuring $X$-type stabilizers. However, the support of any such hook error lies entirely in one column, because the measurement pattern of the $X$-type stabilizers is N-shaped. Thus, we have shown that any faulty component only populates at most a single column with $X$-errors, and therefore at least $d$ are needed to cause $\bar X$. The argument that $\bar Z$ requires at least $d$ faults to construct is essentially identical. Constructing $\bar Y$ requires constructing both a column traversing string of $X$s and a row traversing string of $Z$s. Since $X$ and $Z$ errors are edges on disconnected decoding graphs, $\bar Y$ requires at least $2d-1$ faults.

Now consider syndrome extraction Fig.~\ref{circ_synextract} for the distance-$d$ triangle code. Notice that the planes are simply patches of surface code undergoing syndrome extraction according to Fig.~\ref{circ_surfacecode}, which we know is full-distance. The fact that syndrome measurement is staggered from plane to plane can only make it harder to place an undetected logical error using fewer than $d$ faults (since later measurements may pick up on earlier placed errors). Ignoring the staggered timing, we can still argue for effective distance $d$ of the syndrome extraction. 

The outline of the argument is this: we first argue that $\bar Y$ cannot be created with fewer than $d-1$ faults. By symmetry, $\bar X$ and $\bar Z$ are are similarly resilient and this argues that Fig.~\ref{circ_synextract} has effective distance $d-1$. Then, again appealing to symmetry, we note that the only barrier to achieving effective distance $d$ is the origin. Finally, case by case analysis of hooks near the origin shows that some hooks are indeed dangerous (two hooks can cause a weight three error as noted in Sec.~\ref{sec:init_meas} of the main text) and that using two 2-CATs for syndrome extraction of checks adjacent to the origin is sufficient to detect the dangerous hooks.

To begin, notice that $\bar Y$ must anticommute with any $\bar X$ or $\bar Z$. Each row of qubits with fixed $x$-coordinate $x>0$ supports a string of $X$s acting as $\bar X$. Thus, since it must anticommute, $\bar Y$ has an odd number of $Z$s in each of these rows. Define an asymmetric surface code, call it code $x$, from the qubits in the triangle code with coordinate $x>0$. Notice that Fig.~\ref{circ_synextract} implies that code $x$ undergoes syndrome extraction with effective distance $(d-1)/2$, because, as described in the case of the surface code Fig.~\ref{circ_surfacecode}, the hooks are correctly oriented. Now, $\bar Y$ must act like $\bar Z_x$ (i.e.~logical $Z$ for code $x$) when restricted to code $x$ (that is, when the support of $\bar Y$ not within code $x$ is ignored). Thus, the full-distance syndrome extraction of code $x$ says that we cannot create $\bar Y$ with fewer than $(d-1)/2$ faults.

We can similarly define a surface code from only the qubits at coordinates $z\ge0$, code $z$. By its need to anticommute with any $\bar Z$ string, $\bar Y$ must act like $\bar X_z$ (logical $X$ of code $z$) when restricted to code $z$. Since code $z$ is undergoing full-distance extraction, it takes at least $(d+1)/2$ faults to make $\bar Y$.

We now combine the two observations that $\bar Y$ acts like $\bar Z_x$ and $\bar X_z$ under the appropriate restrictions. By restricting to codes $x$ and $z$, we have disconnected the $X$ and $Z$ decoding graphs of Fig.~\ref{app_decodegraph}. Thus, $\bar Y$ restricted to codes $x$ and $z$ cannot be created with fewer than $(d-1)/2+(d+1)/2-1=d-1$ faults, finishing the argument for effective distance $d-1$.

Now, the location of the mixed-type stabilizers is not fundamental, and local-Clifford equivalent triangle codes exist with it orientated along any ray from the origin. So the argument above could be made with mixed types along any axis instead of the $y$-axis. What is necessary is that the ray of mixed-type stabilizers terminate in a twist at the origin. Checking hooks at the origin, we see that two hook faults can cause the weight-3 error $Z_{(1,0,0)}Y_{(0,0,0)}X_{(0,0,1)}$, which is three of the supporting Paulis of 
\begin{equation}
\bar Y=-i\prod_{x\in\mathbb{Z}_s}Z_{(x,0,0)}\prod_{z\in\mathbb{Z}_s}X_{(0,0,z)},
\end{equation}
with $s=(d+1)/2$. This implies $\bar Y$ can be made from $d-1$ faults. To catch this case, introduce 2-CAT decoding (Fig.~\ref{circ_flags}) on any two of the central plaquettes, so that any two central hooks can be detected.

\section{exREC pseudothresholds}\label{app_pseudothresh}
Here we discuss the fault-tolerant designs and calculations making up Tab.~\ref{d3_smallFTcompare} in the text. All our calculations are rigorous and performed in the exREC formalism \cite{Aliferis2006}. We therefore report pseudothresholds \cite{Svore2006} for \emph{computation} (albeit for just Clifford gates on a single logical qubit), as opposed to thresholds for topological memory, which were discussed earlier in Appendix~\ref{app_thresh}. Since such pseudothreshold results are not asymptotic, we consider them more suited to comparing small experiments. Eventually one may want to scale up the calculation to simulate the error rates of small algorithms (such as is considered in \cite{Gottesman2016}) made from several exRECs.

\begin{figure}
\includegraphics[width=\columnwidth]{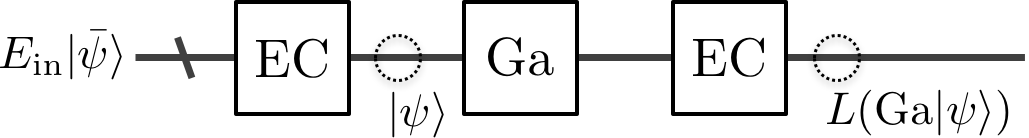}
\caption{\label{app_exREC} An exREC \cite{Aliferis2006} consists of a logical gate gadget Ga sandwiched between two error-correction gadgets EC. The gates we consider are transversal, so Ga is simply a depth-1 circuit of single-qubit gates. The EC gadgets are more complex and described in the text. Ideal decoding (dotted circles) after the first and second ECs allows one to determine the Pauli operator $L$, which depends on faults that have occurred in the exREC, but not, if the CDT criteria \cite{Cross2007} are satisfied, on incoming errors $E_{\text{in}}$ (here assumed, without loss of generality, to contain no logical errors). If $L$ is the identity, no fault has occurred, but $X$, $Y$, or $Z$ signals failure of the exREC. Counting the number of ways failure occurs from $\le2$ faults gives our pseudothreshold bounds.}
\end{figure}

In the exREC formalism it is important to create an error-correction (EC) circuit that can recover from any $(d-1)/2$ faults in the preceeding gate (Ga) or the preceeding EC. In particular, for distance-3, any single fault in the so-called exREC circuit EC.Ga.EC must not lead to a logical error. A logical error in turn is defined as when an ideal decoding of the state after the trailing EC differs from the expected state $\text{Ga}\ket{\psi}$ given that an ideal decoding after the leading EC revealed state $\ket{\psi}$ (see Fig.~\ref{app_exREC}).

All our EC gadgets for distance-3 codes are constructed in the same way. First, define a syndrome extraction (SE) circuit unique to a given design. Then perform $\text{SE}.\text{SE}$ obtaining two copies of the syndrome. If these syndromes match, the EC gadget is complete, and we attempt to deduce an appropriate recovery (a Pauli operator). If the syndromes do not match, we repeat SE once more, and then attempt recovery. The recovery is never explicitly applied; we simply adjust the Pauli frame \cite{Knill2005}.

We have to specify how a recovery operator is deduced, often called ``decoding'' the syndrome. Because we are dealing with finite-sized codes designed specifically for a small demonstration of fault-tolerance, we advocate table lookup for maximum pseudothreshold. The construction of the table proceeds in two steps. First, EC must be able to correct all errors arising from a single fault that it introduces itself, either immediately or, if the error gets through, in the next round of EC. We can enumerate these cases by brute force simulation of the Clifford exREC circuit, and add them to the table. If ever there are two or more errors with the same syndrome but which commute differently with the logical operators, then we know the SE is not fault-tolerant and must redesign it. The second step of filling the table is to take unused syndromes and assign to each the lowest weight Pauli recovery that corrects the \emph{latest} syndrome measured. Although these syndrome patterns did not arise from single faults, they may arise from double faults, and so we can correct some double faults. However, we make no attempt to correct the most likely (assuming circuit depolarizing noise) double faults consistent with a measured syndrome.

There are two properties of the EC constructed this way that are important. First, it is fault-tolerant in the sense mentioned above that any single fault in the exREC does not cause a logical error. Second, it obeys the Cross-DiVincenzo-Terhal (CDT) criteria \cite{Cross2007}. This guarantees that errors incoming to the exREC will not affect the logical failure rate, and thus we can ignore them when calculating pseudothresholds, rather than performing, for instance, an analysis in the context of a worst-case input.

As a detail, we note that $\text{SE}.\text{SE}$ can sometimes be simplified by further parallelization of the circuit. For instance, Fig.~\ref{circ_trianglecode3} is one such case, where the depth of $\text{SE}.\text{SE}$ is one less than twice the depth of SE. We perform such simplifications when possible, but do not endeavor to simplify $\text{SE}.\text{SE}.\text{SE}$ if it occurs. This is because in general it cannot be decided whether the third SE should be applied before the second has completely finished.

Having laid out the general procedure for building the EC gadget, we need to now specify SE for each design. We present one SE for each code, and mention the variants, which are straightforward. Afterward, we discuss the counting of malignant faults necessary to actually rigorously bound the pseudothresholds.

\begin{figure}
\hspace*{0.15in}
\includegraphics[width=0.85\columnwidth]{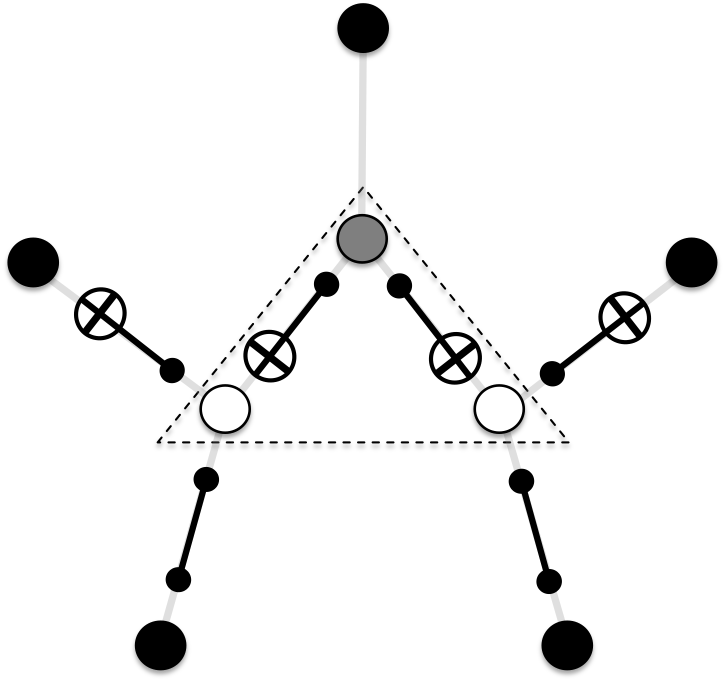}
\caption{\label{5qub_layout} A layout of the 5-qubit code (black circles) and ancilla qubits (gray, prepared as $\ket{+}$, and white, prepared as $\ket{0}$) overlaid with the circuit for measurement of a single stabilizer $XZZXI$. The gates inside the dotted triangle are performed twice, both before and after the gates outside the triangle. For fault-tolerance the order of all gates is otherwise irrelevant, though it will change the decoding table. All ancilla qubits are measured afterward in the same basis they were prepared. Light lines in the background indicate the required qubit interaction graph for an entire SE cycle, which is planar and only degree three.}
\end{figure}

Syndrome extraction on the 5-qubit code \cite{Bennett1996,Laflamme1996}
is typically done with verified 4-CAT states. However, we find verification unsuitable for a small fault-tolerant design, since it requires postselection. In principle, the verification can be removed by decoding the CAT states before measurement \cite{Divincenzo2007}. But if we are going to use decoding, there is a more compact strategy using just 3-CATs. In fact, just one set of three ancillas suffices, as shown in Fig.~\ref{5qub_layout}, since they can be prepared repeatedly in a 3-CAT state to measure all stabilizers. This is notable as it is the smallest completely fault-tolerant design that we know of, using just 8 qubits including ancillas. Variations on this design using more ancillas are possible, such as using two sets of three ancillas (which is still planar) and four sets of three (which is not), to increase parallelism of the syndrome extraction and correspondingly the threshold.

\begin{figure}
\hspace*{0.25in}
\includegraphics[width=0.8\columnwidth]{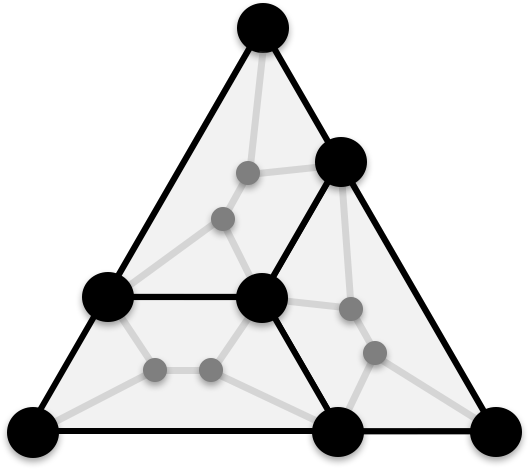}
\caption{\label{7qub_layout} A 6-ancilla design for the smallest color code. With only degree-3 connectivity (light-gray edges), this layout provides fully fault-tolerant syndrome extraction by creating and decoding 2-CATs.}
\end{figure}

For the smallest color code, equivalent to Steane's code \cite{Steane1996}, it has been suggested that 2-CATs are sufficient for full-distance syndrome extraction \cite{Stephens2014,Landahl2014}. Indeed we find this to be the case, \emph{if} we decode the 2-CATs to detect hook errors. In fact, if we did not decode, the high symmetry of the 7-qubit code renders any undetected hook error fatal. The smallest fault-tolerant planar design is shown in Fig.~\ref{7qub_layout} where each face possesses two ancillas. These must be prepared and reprepared in 2-CAT states to measure all stabilizers (since each face represents both an $X$-type and $Z$-type stabilizer). If instead we introduce four ancillas per face such that the $X$- and $Z$-syndromes might be extracted more in parallel (e.g.~the ancillas for $Z$-type can now be prepared while those for $X$-type are being measured), we lose planarity but increase the threshold. 

\begin{figure}[t]
\hspace*{0.25in}
\includegraphics[width=0.8\columnwidth]{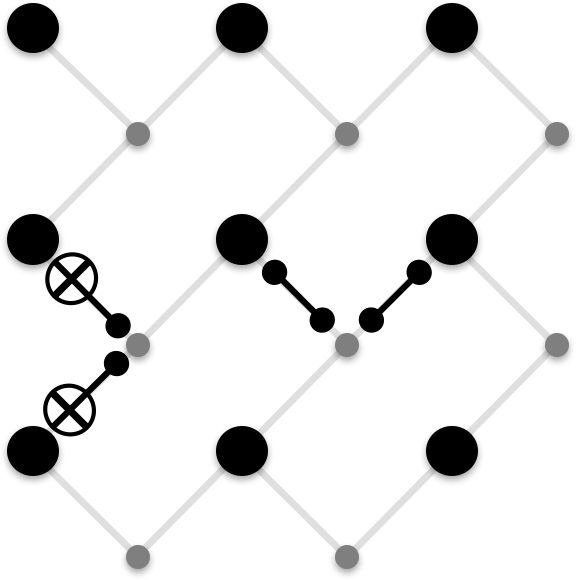}
\caption{\label{9qub_layout} The $3\times3$ Bacon-Shor code with 8 ancilla qubits. Overlaid are the coupling gates required to measure an $X$-type gauge operator (to the left) and a $Z$-type gauge operator (right) when the ancillas are prepared and measured in the $X$-basis. Light-gray lines indicate the degree-3 connectivity required for full syndrome extraction.}
\end{figure}

The Bacon-Shor code is unique in our study as it is a subsystem code. When syndrome extraction is performed with ancilla codeblocks, its threshold is one of the highest known \cite{Aliferis2007}. But as pointed out in \cite{Aliferis2007}, we can exploit the subsystem structure to measure only weight-two operators, also called gauge operators, to deduce the syndrome. Since weight-two measurements cannot introduce more than weight-one errors to the data, this procedure is naturally fault-tolerant. To minimize ancilla reuse, while keeping under 20 total qubits (our cutoff for ``small" fault-tolerant designs), we use a total of eight ancillas, Fig.~\ref{9qub_layout}. SE consists of measuring all $X$-type gauge operators once, followed by measurement of all the $Z$-type gauge operators.

We have described several SE designs for the triangle code in Section~\ref{sec:distance_3}. The three in Tab.~\ref{d3_smallFTcompare} are (1) the two 2-CAT design suggested by the distance-3 version of Fig.~\ref{circ_synextract}, (2) the degree-5 version of Fig.~\ref{circ_trianglecode3} wherein plaquette ancillas can interact directly with neighboring loop ancillas, and (3) the degree-4 version of Fig.~\ref{circ_trianglecode3} wherein the ancillas interact only via a data qubit intermediary.

The surface code SE is taken from \cite{Tomita2014}. There they conclude that the 8-ancilla version possesses the highest threshold. We also consider the 4-ancilla version, in which plaquette ancillas are reused to measure loops, because it uses 13 total qubits and so offers an equal-qubit comparison to our smallest triangle code design. 


Once we have completely specified the exREC for a design, we can rigorously bound the logical error rate. Our error model is standard. Single-qubit operations including identity gates fail with $X$, $Y$, or $Z$ with probability $p/3$ while two-qubit gates fail in one of 15 Pauli ways with probability $p/15$ each. Single-qubit preparation of $\ket{0}$ or $\ket{+}$ fails (by a Pauli error $X$ or $Z$, respectively) with probability $p$. Measurement in either the $X$- or $Z$-basis fail (by reporting the wrong bit) with probability $p$. All circuit components succeed or fail independently.

Counting malignant sets proceeds similarly to \cite{Aliferis2006}, though our bounds are calculated using different (slightly tighter) formulas. We can calculate the quantities
\begin{align}
P_{\text{fail}}^{(2)}&=\text{Pr}\left[L\neq I\text{, }\le2\text{ faults}\right],\\
P_{\text{succ}}^{(2)}&=\text{Pr}\left[L=I\text{, }\le2\text{ faults}\right]
\end{align}
(where $I$ is the $2\times2$ identity) by enumerating all sets of at most two faults and determining $L$ (see Fig.~\ref{app_exREC}) for each. The probability that a particular set of faults occurs is the product of the probabilities for the faulty components failing and for all other components succeeding.

Defining $P_{\text{fail}}=\text{Pr}\left[L\neq I\right]$, we then see the bounds,
\begin{align}
P_{\text{fail}}^{(2)}\le P_{\text{fail}}\le 1-P_{\text{succ}}^{(2)}.
\end{align}
By the fault-tolerance of our designs, $P_{\text{fail}}=O(p^2)$, and likewise with the upper and lower bounds. So the pseudothreshold $p_{\text{th}}$ found by solving $P_{\text{fail}}(p_{\text{th}})=p_{\text{th}}$ is bounded by the solutions to
\begin{align}
P_{\text{fail}}^{(2)}(p_{\text{upp}})&=p_{\text{upp}},\\
1-P_{\text{succ}}^{(2)}(p_{\text{low}})&=p_{\text{low}}
\end{align}
like  $p_{\text{low}}\le p_{\text{th}}\le p_{\text{upp}}$.

Note that counting sets of faults is complicated by the fact our EC is non-deterministic (the third application of SE is conditional on the first two not matching). Thus, to calculate $P_{\text{fail}}^{(2)}$ and $P_{\text{succ}}^{(2)}$ we must break each into four cases. This amounts to the identities
\begin{align}
P_{\text{fail}}^{(2)}&=\hspace{-5pt}\sum_{i,j\in\{2,3\}}\hspace{-5pt}\text{Pr}\left[L\neq I,\le2\text{ faults},i\text{-EC}_1,j\text{-EC}_2\right]\\
P_{\text{succ}}^{(2)}&=\hspace{-5pt}\sum_{i,j\in\{2,3\}}\hspace{-5pt}\text{Pr}\left[L=I,\le2\text{ faults},i\text{-EC}_1,j\text{-EC}_2\right],
\end{align}
where $i\text{-EC}_1$ indicates that the first EC consists of two ($i=2$) or three ($i=3$) SE, and likewise with $j\text{-EC}_2$ for the second EC. The probabilities in the sums can be calculated by simulating all sets of at most two faults in a circuit with an $i\text{-EC}_1$ and a $j\text{-EC}_2$ and then keeping only the sets of faults which cause syndromes that are consistent with the presence of an $i\text{-EC}_1$ and a $j\text{-EC}_2$.

\begin{figure}
\includegraphics[width=0.9\columnwidth]{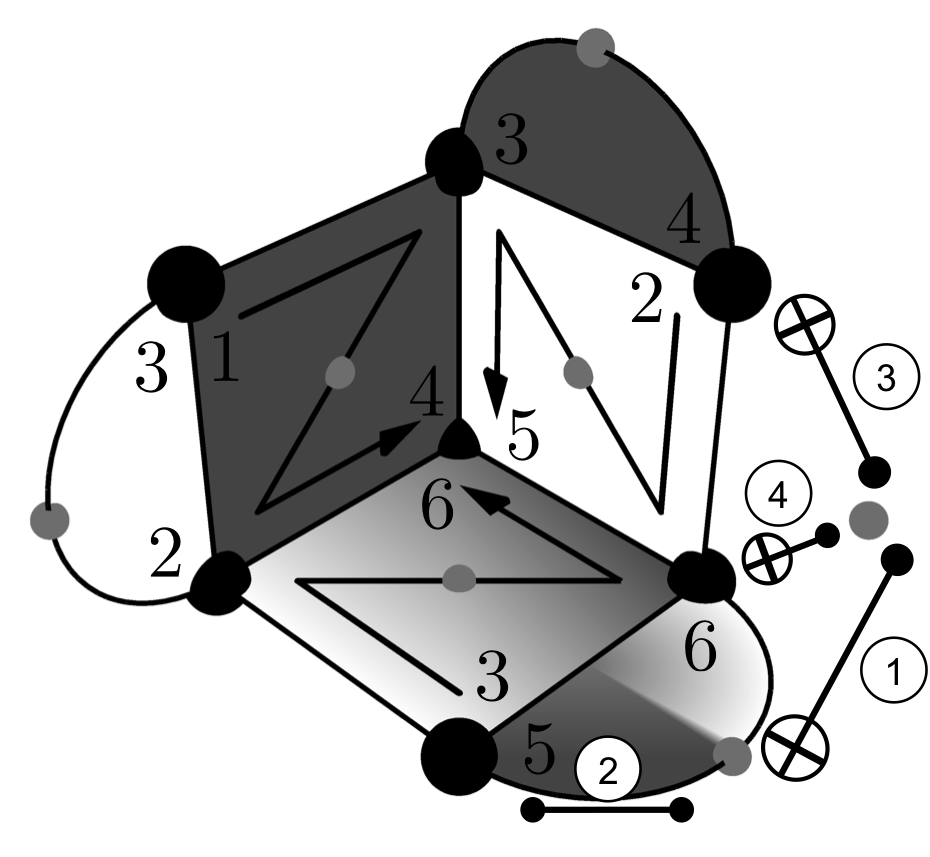}
\caption{\label{app_measplusone} Measuring $\bar Y$ using one extra qubit beyond Fig.~\ref{circ_trianglecode3}. Gates are labeled with the timesteps in which they should be applied. The extra qubit is prepared (during timestep zero) in $\ket{+}$ and the loop ancilla for the mixed-type loop is prepared in $\ket{0}$ so that the first CNOT creates a 2-CAT state for measuring $\bar Y$. This allows the loop ancilla to be measured and reprepared in $\ket{+}$ to measure the loop stabilizer in accordance with the scheduling of Fig.~\ref{circ_trianglecode3}. Thus, measuring $\bar Y$ and all the stabilizers takes no more timesteps than measuring just the stabilizers.}
\end{figure}

Having discussed gate exRECs, we briefly describe the procedure for calculating the error rate of measurement exRECs, which we performed for the small triangle code designs in Tab.~\ref{d3_meascompare}. A measurement exREC is set up similarly to Fig.~\ref{app_exREC}, except that the second EC is replaced by a measurement EC, or MEC. Thus, the measurement exREC includes both the final single-qubit gate in the circuit and the measurement.

The MEC also consists of repeated syndrome extraction, but one that measures the eigenvalue of a logical Pauli $\bar P$ in addition to the syndromes of all stabilizers. Call this extraction an MSE. Then, an MEC consists of two repeats of the MSE, which is followed by another MSE if the first two did not match. The decoding table of the MSE is constructed just as before for the SE, but, because the initial logical state is unknown, the $\bar P$ values measured cannot provide information about the error directly --- only when they change across different MSEs is it relevant.

The MSE can in principle be done in many different ways. For instance, the procedure described in Section~\ref{sec:distance_3} reuses the plaquette ancilla in the $xy$-plane to do the $\bar P=\bar Z$ measurement. In Fig.~\ref{app_measplusone}, we show how a measurement with one extra ancilla qubit can be done. The counting of malignant sets also proceeds similarly as the gate case, except that final logical errors of $\bar P$ are not considered failures.

\bibliographystyle{unsrtnat}
\bibliography{References}

\begin{thebibliography}{45}
\providecommand{\natexlab}[1]{#1}
\providecommand{\url}[1]{\texttt{#1}}
\expandafter\ifx\csname urlstyle\endcsname\relax
  \providecommand{\doi}[1]{doi: #1}\else
  \providecommand{\doi}{doi: \begingroup \urlstyle{rm}\Url}\fi

\bibitem[Bravyi and Kitaev(1998)]{Bravyi1998}
Sergey~B Bravyi and A~Yu Kitaev.
\newblock Quantum codes on a lattice with boundary.
\newblock \emph{quant-ph/9811052}, 1998.

\bibitem[Dennis et~al.(2002)Dennis, Kitaev, Landahl, and Preskill]{Dennis2002}
Eric Dennis, Alexei Kitaev, Andrew Landahl, and John Preskill.
\newblock Topological quantum memory.
\newblock \emph{Journal of Mathematical Physics}, 43\penalty0 (9):\penalty0
  4452--4505, 2002.
\newblock \doi{10.1063/1.1499754}.

\bibitem[Fowler et~al.(2012)Fowler, Mariantoni, Martinis, and
  Cleland]{Fowler2012c}
Austin~G Fowler, Matteo Mariantoni, John~M Martinis, and Andrew~N Cleland.
\newblock Surface codes: Towards practical large-scale quantum computation.
\newblock \emph{Physical Review A}, 86\penalty0 (3):\penalty0 032324, 2012.
\newblock \doi{10.1103/PhysRevA.86.032324}.

\bibitem[Aharonov and Eldar(2011)]{Aharonov2011}
Dorit Aharonov and Lior Eldar.
\newblock On the complexity of commuting local {Hamiltonians}, and tight
  conditions for topological order in such systems.
\newblock In \emph{Foundations of Computer Science (FOCS), 2011 IEEE 52nd
  Annual Symposium on}, pages 334--343. IEEE, 2011.
\newblock \doi{10.1109/FOCS.2011.58}.

\bibitem[Tomita and Svore(2014)]{Tomita2014}
Yu~Tomita and Krysta~M Svore.
\newblock Low-distance surface codes under realistic quantum noise.
\newblock \emph{Physical Review A}, 90\penalty0 (6):\penalty0 062320, 2014.
\newblock \doi{10.1103/PhysRevA.90.062320}.

\bibitem[Wang et~al.(2011)Wang, Fowler, and Hollenberg]{Wang2011}
David~S Wang, Austin~G Fowler, and Lloyd~CL Hollenberg.
\newblock Surface code quantum computing with error rates over 1\%.
\newblock \emph{Physical Review A}, 83\penalty0 (2):\penalty0 020302, 2011.
\newblock \doi{10.1103/PhysRevA.83.020302}.

\bibitem[Hutter et~al.(2014)Hutter, Wootton, and Loss]{Hutter2014}
Adrian Hutter, James~R Wootton, and Daniel Loss.
\newblock Efficient {Markov} chain {Monte Carlo} algorithm for the surface
  code.
\newblock \emph{Physical Review A}, 89\penalty0 (2):\penalty0 022326, 2014.
\newblock \doi{10.1103/PhysRevA.89.022326}.

\bibitem[Bravyi et~al.(2014)Bravyi, Suchara, and Vargo]{Bravyi2014}
Sergey Bravyi, Martin Suchara, and Alexander Vargo.
\newblock Efficient algorithms for maximum likelihood decoding in the surface
  code.
\newblock \emph{Physical Review A}, 90\penalty0 (3):\penalty0 032326, 2014.
\newblock \doi{10.1103/PhysRevA.90.032326}.

\bibitem[Wootton and Loss(2012)]{Wootton2012}
James~R Wootton and Daniel Loss.
\newblock High threshold error correction for the surface code.
\newblock \emph{Physical Review Letters}, 109\penalty0 (16):\penalty0 160503,
  2012.
\newblock \doi{10.1103/PhysRevLett.109.160503}.

\bibitem[Fowler and Devitt(2012)]{Fowler2012a}
Austin~G Fowler and Simon~J Devitt.
\newblock A bridge to lower overhead quantum computation.
\newblock \emph{arXiv:1209.0510}, 2012.

\bibitem[Bomb{\'\i}n(2010{\natexlab{a}})]{Bombin2010a}
H{\'e}ctor Bomb{\'\i}n.
\newblock Topological subsystem codes.
\newblock \emph{Physical Review A}, 81\penalty0 (3):\penalty0 032301,
  2010{\natexlab{a}}.
\newblock \doi{10.1103/PhysRevA.81.032301}.

\bibitem[Bravyi et~al.(2013)Bravyi, Duclos-Cianci, Poulin, and
  Suchara]{Bravyi2012}
Sergey Bravyi, Guillaume Duclos-Cianci, David Poulin, and Martin Suchara.
\newblock Subsystem surface codes with three-qubit check operators.
\newblock \emph{Quantum Information \& Computation}, 13\penalty0
  (11-12):\penalty0 963--985, 2013.

\bibitem[Bomb{\'\i}n and Martin-Delgado(2006)]{Bombin2006}
H{\'e}ctor Bomb{\'\i}n and Miguel~Angel Martin-Delgado.
\newblock Topological quantum distillation.
\newblock \emph{Physical Review Letters}, 97\penalty0 (18):\penalty0 180501,
  2006.
\newblock \doi{10.1103/PhysRevLett.97.180501}.

\bibitem[Bomb{\'\i}n(2015)]{Bombin2015}
H{\'e}ctor Bomb{\'\i}n.
\newblock Gauge color codes: optimal transversal gates and gauge fixing in
  topological stabilizer codes.
\newblock \emph{New Journal of Physics}, 17\penalty0 (8):\penalty0 083002,
  2015.
\newblock \doi{10.1088/1367-2630/17/8/083002}.

\bibitem[Kubica and Beverland(2015)]{Kubica2015a}
Aleksander Kubica and Michael~E Beverland.
\newblock Universal transversal gates with color codes: A simplified approach.
\newblock \emph{Physical Review A}, 91\penalty0 (3):\penalty0 032330, 2015.
\newblock \doi{10.1103/PhysRevA.91.032330}.

\bibitem[Landahl and Ryan-Anderson(2014)]{Landahl2014}
Andrew~J Landahl and Ciaran Ryan-Anderson.
\newblock Quantum computing by color-code lattice surgery.
\newblock \emph{arXiv:1407.5103}, 2014.

\bibitem[Raussendorf and Harrington(2007)]{Raussendorf2007}
Robert Raussendorf and Jim Harrington.
\newblock Fault-tolerant quantum computation with high threshold in two
  dimensions.
\newblock \emph{Physical Review Letters}, 98\penalty0 (19):\penalty0 190504,
  2007.
\newblock \doi{10.1103/PhysRevLett.98.190504}.

\bibitem[Bombin and Martin-Delgado(2009)]{Bombin2009}
H~Bombin and MA~Martin-Delgado.
\newblock Quantum measurements and gates by code deformation.
\newblock \emph{Journal of Physics A: Mathematical and Theoretical},
  42\penalty0 (9):\penalty0 095302, 2009.
\newblock \doi{10.1088/1751-8113/42/9/095302}.

\bibitem[Bravyi and Kitaev(2005)]{Bravyi2005}
Sergey Bravyi and Alexei Kitaev.
\newblock Universal quantum computation with ideal clifford gates and noisy
  ancillas.
\newblock \emph{Physical Review A}, 71\penalty0 (2):\penalty0 022316, 2005.
\newblock \doi{10.1103/PhysRevA.71.022316}.

\bibitem[Fowler et~al.(2013)Fowler, Devitt, and Jones]{Fowler2013}
Austin~G Fowler, Simon~J Devitt, and Cody Jones.
\newblock Surface code implementation of block code state distillation.
\newblock \emph{Scientific reports}, 3, 2013.
\newblock \doi{10.1038/srep01939}.

\bibitem[Fowler et~al.(2009)Fowler, Stephens, and Groszkowski]{Fowler2009}
Austin~G Fowler, Ashley~M Stephens, and Peter Groszkowski.
\newblock High-threshold universal quantum computation on the surface code.
\newblock \emph{Physical Review A}, 80\penalty0 (5):\penalty0 052312, 2009.
\newblock \doi{10.1103/PhysRevA.80.052312}.

\bibitem[Fowler(2012)]{Fowler2012b}
Austin~G Fowler.
\newblock Time-optimal quantum computation.
\newblock \emph{arXiv:1210.4626}, 2012.

\bibitem[Horsman et~al.(2012)Horsman, Fowler, Devitt, and
  Van~Meter]{Horsman2012}
Clare Horsman, Austin~G Fowler, Simon Devitt, and Rodney Van~Meter.
\newblock Surface code quantum computing by lattice surgery.
\newblock \emph{New Journal of Physics}, 14\penalty0 (12):\penalty0 123011,
  2012.
\newblock \doi{10.1088/1367-2630/14/12/123011}.

\bibitem[Bomb{\'\i}n(2010{\natexlab{b}})]{Bombin2010b}
H{\'e}ctor Bomb{\'\i}n.
\newblock Topological order with a twist: {Ising} anyons from an {Abelian}
  model.
\newblock \emph{Physical Review Letters}, 105\penalty0 (3):\penalty0 030403,
  2010{\natexlab{b}}.
\newblock \doi{10.1103/PhysRevLett.105.030403}.

\bibitem[Hastings and Geller(2015)]{Hastings2014}
Matthew~B Hastings and A~Geller.
\newblock Reduced space-time and time costs using dislocation codes and
  arbitrary ancillas.
\newblock \emph{Quantum Information \& Computation}, 15\penalty0
  (11-12):\penalty0 962--986, 2015.

\bibitem[Calderbank and Shor(1996)]{Calderbank1996}
A.~R. Calderbank and Peter~W. Shor.
\newblock Good quantum error-correcting codes exist.
\newblock \emph{Phys. Rev. A}, 54:\penalty0 1098--1105, 1996.
\newblock \doi{10.1103/PhysRevA.54.1098}.

\bibitem[Steane(1996)]{Steane1996}
Andrew~M Steane.
\newblock Error correcting codes in quantum theory.
\newblock \emph{Phys. Rev. Lett.}, 77\penalty0 (5):\penalty0 793, 1996.
\newblock \doi{10.1103/PhysRevLett.77.793}.

\bibitem[Brown et~al.(2016)Brown, Laubscher, Kesselring, and
  Wootton]{Brown2016}
Benjamin~J Brown, Katharina Laubscher, Markus~S Kesselring, and James~R
  Wootton.
\newblock Poking holes and cutting corners to achieve {Clifford} gates with the
  surface code.
\newblock \emph{arXiv:1609.04673}, 2016.

\bibitem[Bombin and Martin-Delgado(2007)]{Bombin2007}
H~Bombin and MA~Martin-Delgado.
\newblock Optimal resources for topological two-dimensional stabilizer codes:
  Comparative study.
\newblock \emph{Physical Review A}, 76\penalty0 (1):\penalty0 012305, 2007.
\newblock \doi{10.1103/PhysRevA.76.012305}.

\bibitem[Gottesman and Chuang(1999)]{Gottesman1999}
Daniel Gottesman and Isaac~L Chuang.
\newblock Demonstrating the viability of universal quantum computation using
  teleportation and single-qubit operations.
\newblock \emph{Nature}, 402\penalty0 (6760):\penalty0 390--393, 1999.
\newblock \doi{10.1038/46503}.

\bibitem[Shor(1996)]{Shor1996}
Peter~W Shor.
\newblock Fault-tolerant quantum computation.
\newblock In \emph{Foundations of Computer Science, 1996. Proceedings., 37th
  Annual Symposium on}, pages 56--65. IEEE, 1996.
\newblock \doi{10.1109/SFCS.1996.548464}.

\bibitem[Brooks and Preskill(2013)]{Brooks2013}
Peter Brooks and John Preskill.
\newblock Fault-tolerant quantum computation with asymmetric {Bacon-Shor}
  codes.
\newblock \emph{Physical Review A}, 87\penalty0 (3):\penalty0 032310, 2013.
\newblock \doi{10.1103/PhysRevA.87.032310}.

\bibitem[Moussa(2016)]{Moussa2016}
Jonathan~E Moussa.
\newblock Transversal clifford gates on folded surface codes.
\newblock \emph{Physical Review A}, 94\penalty0 (4):\penalty0 042316, 2016.
\newblock \doi{10.1103/PhysRevA.94.042316}.

\bibitem[Kubica et~al.(2015)Kubica, Yoshida, and Pastawski]{Kubica2015b}
Aleksander Kubica, Beni Yoshida, and Fernando Pastawski.
\newblock Unfolding the color code.
\newblock \emph{New Journal of Physics}, 17\penalty0 (8):\penalty0 083026,
  2015.
\newblock \doi{10.1088/1367-2630/17/8/083026}.

\bibitem[Cross()]{Cross2016}
Andrew~W Cross.
\newblock personal communication.

\bibitem[Aliferis et~al.(2006)Aliferis, Gottesman, and Preskill]{Aliferis2006}
Panos Aliferis, Daniel Gottesman, and John Preskill.
\newblock Quantum accuracy threshold for concatenated distance-3 codes.
\newblock \emph{Quantum Information \& Computation}, 6\penalty0 (2):\penalty0
  97--165, 2006.

\bibitem[Svore et~al.(2006)Svore, Cross, Chuang, and Aho]{Svore2006}
Krysta~M Svore, Andrew~W Cross, Isaac~L Chuang, and Alfred~V Aho.
\newblock A flow-map model for analyzing pseudothresholds in fault-tolerant
  quantum computing.
\newblock \emph{Quantum Information \& Computation}, 6\penalty0 (3):\penalty0
  193--212, 2006.

\bibitem[Gottesman(2016)]{Gottesman2016}
Daniel Gottesman.
\newblock Quantum fault tolerance in small experiments.
\newblock \emph{arXiv preprint arXiv:1610.03507}, 2016.

\bibitem[Cross et~al.(2009)Cross, Divincenzo, and Terhal]{Cross2007}
Andrew~W Cross, David~P Divincenzo, and Barbara~M Terhal.
\newblock A comparative code study for quantum fault tolerance.
\newblock \emph{Quantum Information \& Computation}, 9\penalty0 (7):\penalty0
  541--572, 2009.

\bibitem[Knill(2005)]{Knill2005}
Emanuel Knill.
\newblock Quantum computing with realistically noisy devices.
\newblock \emph{Nature}, 434\penalty0 (7029):\penalty0 39--44, 2005.
\newblock \doi{10.1038/nature03350}.

\bibitem[Bennett et~al.(1996)Bennett, DiVincenzo, Smolin, and
  Wootters]{Bennett1996}
Charles~H Bennett, David~P DiVincenzo, John~A Smolin, and William~K Wootters.
\newblock Mixed-state entanglement and quantum error correction.
\newblock \emph{Physical Review A}, 54\penalty0 (5):\penalty0 3824, 1996.
\newblock \doi{10.1103/PhysRevA.54.3824}.

\bibitem[Laflamme et~al.(1996)Laflamme, Miquel, Paz, and Zurek]{Laflamme1996}
Raymond Laflamme, Cesar Miquel, Juan~Pablo Paz, and Wojciech~Hubert Zurek.
\newblock Perfect quantum error correcting code.
\newblock \emph{Physical Review Letters}, 77\penalty0 (1):\penalty0 198, 1996.
\newblock \doi{10.1103/PhysRevLett.77.198}.

\bibitem[DiVincenzo and Aliferis(2007)]{Divincenzo2007}
David~P DiVincenzo and Panos Aliferis.
\newblock Effective fault-tolerant quantum computation with slow measurements.
\newblock \emph{Physical Review Letters}, 98\penalty0 (2):\penalty0 020501,
  2007.
\newblock \doi{10.1103/PhysRevLett.98.020501}.

\bibitem[Stephens(2014)]{Stephens2014}
Ashley~M Stephens.
\newblock Efficient fault-tolerant decoding of topological color codes.
\newblock \emph{arXiv preprint arXiv:1402.3037}, 2014.

\bibitem[Aliferis and Cross(2007)]{Aliferis2007}
Panos Aliferis and Andrew~W Cross.
\newblock Subsystem fault tolerance with the {Bacon-Shor} code.
\newblock \emph{Physical Review Letters}, 98\penalty0 (22):\penalty0 220502,
  2007.
\newblock \doi{10.1103/PhysRevLett.98.220502}.

\end{thebibliography}

\end{document}